\documentclass[preprint, 12pt, onecolumn, superscriptaddress, frontmatterverbose, amsmath,amssymb]{revtex4-1}

\usepackage{gensymb}
\usepackage{amsmath}
\usepackage{nccmath}
\usepackage{graphics}
\usepackage{epsfig}
\usepackage{bm}
\usepackage{dcolumn}
\usepackage{float}
\usepackage{nicefrac}
\renewcommand{\vec}[1]{\mathbf{#1}}

\begin{document}

\title{Nuclear Spin Assisted Magnetic Field Angle Sensing}

\author{Ziwei Qiu}
\affiliation{Department of Physics, Harvard University, Cambridge, MA 02138, USA}
\affiliation{John A. Paulson School of Engineering and Applied Sciences, Harvard University, Cambridge, MA 02138, USA}
\author{Uri Vool}
\affiliation{Department of Physics, Harvard University, Cambridge, MA 02138, USA}
\affiliation{John Harvard Distinguished Science Fellows Program, Harvard University, Cambridge MA 02138, USA}
\author{Assaf Hamo}
\affiliation{Department of Physics, Harvard University, Cambridge, MA 02138, USA}
\author{Amir Yacoby}
\email[Correspondence to: yacoby@physics.harvard.edu]{}
\affiliation{Department of Physics, Harvard University, Cambridge, MA 02138, USA}
\affiliation{John A. Paulson School of Engineering and Applied Sciences, Harvard University, Cambridge, MA 02138, USA}

\begin{abstract}

Quantum sensing exploits the strong sensitivity of quantum systems to measure small external signals. The nitrogen-vacancy (NV) center in diamond is one of the most promising platforms for real-world quantum sensing applications, predominantly used as a magnetometer. However, its magnetic field sensitivity vanishes when a bias magnetic field acts perpendicular to the NV axis. Here, we introduce a novel sensing strategy assisted by the nitrogen nuclear spin that uses the entanglement between the electron and nuclear spins to restore the magnetic field sensitivity. This, in turn, allows us to detect small changes in the magnetic field angle relative to the NV axis. Furthermore, based on the same underlying principle, we show that the NV coupling strength to magnetic noise, and hence its coherence time, exhibits a strong asymmetric angle dependence. This allows us to uncover the directional properties of the local magnetic environment and to realize maximal decoupling from anisotropic noise.

\end{abstract}

\maketitle

\section*{Introduction}
Quantum sensing harnesses the coherence of well-controlled quantum systems to detect small signals with high sensitivity \cite{budker2007optical, taylor2008high, Degen2017}. Typically, an external signal directly leads to a shift of the quantum sensor's energy levels. Ancillary sensors, which do not interact with the signal directly, can assist the main sensor by, for example, acting as a long-lived quantum memory \cite{dutt2007quantum, Jiang2009, zaiser2016enhancing, maurer2012room, zaiser2016enhancing, rosskopf2017quantum} or providing error correction \cite{dur2014improved, arrad2014increasing, kessler2014quantum, taminiau2014universal, hirose2016coherent, unden2016quantum}. Solid state spins are promising platforms for quantum sensing techniques and applications, among which nitrogen-vacancy (NV) centers in diamond have received the most attention \cite{taylor2008high, schirhagl2014nitrogen, casola2018probing, bucher2019quantum}. The electron spin associated with the negatively-charged NV center has long coherence time even at room temperature and is capable of detecting a variety of signals with high sensitivity and nanoscale resolution. These include magnetic \cite{maze2008nanoscale, mamin2013nanoscale, hong2013nanoscale} and electric fields \cite{Dolde2011, Michl2019, mittiga2018imaging, iwasaki2017direct, block2020optically, PhysRevLett.124.247701}, temperature \cite{kucsko2013nanometre, neumann2013high, toyli2013fluorescence, Choi201922730}, and pressure \cite{doherty2014electronic, ivady2014pressure, kehayias2019imaging}. 

For the NV to act as a magnetometer, a bias magnetic field along the NV axis is generally required to put the electron spin (S=1) in the $|m_S=0, \pm 1\rangle$ basis, such that the energy levels are first-order sensitive to magnetic field perturbations \cite{taylor2008high}. However, this method fails when the bias field turns toward the direction perpendicular to the NV axis, where the Zeeman interaction no longer induces energy shifts between the levels. To unlock the full potential under this unfavored condition, we introduce an entirely different sensing approach assisted by the ancillary $^{15}$N nuclear spin.  The essential principle is based on the sensitivity of the hyperfine interaction (between the electron and nuclear spins) to small magnetic signals. By monitoring the entanglement between the two spins using spin-echo sequences, we detect small changes in the magnetic field angle. Furthermore, similar to the hyperfine interaction, we show that the coupling between the electron spin and magnetic noise also sensitively depends on the bias field angle, which can be further employed to distinguish and characterize anisotropic noise in the environment. Our exploration extends the capabilities of the versatile sensing toolkit of the NV center, and this new sensing strategy, based on the interaction between the main sensor and ancillary sensor, can be implemented on other quantum sensing platforms as well.

Our experiments were done under ambient conditions, as schematically depicted in Fig. 1a-b. The single-crystal diamond chip contains individually resolvable $^{15}$NVs. A metal stripline fabricated on the diamond surface delivers microwave currents to manipulate the NV spin states. A cylindrical permanent magnet exerts a magnetic field at the NV and provides a coarse control of the field angle ($\theta_B$), while a small DC current flowing in the stripline fine-tunes $\theta_B$ in the close vicinity of $90\degree$. We define $\hat{z}$ to be the NV axis and $\hat{x}$ to be the bias magnetic field direction when it is exactly perpendicular (Fig. 1a).

The paper is structured as follows. We begin by analyzing the NV electron energy eigenstates as the magnetic field angle $\theta_B$ varies around $90\degree$. In particular, we study how the angle modulates the electron spin operator expectation values $\langle \vec{S} \rangle$ at each state, and consequently affects its interaction with the $^{15}$N nuclear spin. Next, we demonstrate magnetic field angle sensing by using spin-echo interferometry to measure the angle-dependent hyperfine interaction. Lastly, we show that the NV coherence exhibits an asymmetric angle dependence, which originates from anisotropic noise in the environment.

\section*{Results}
\subsection*{Electron spin eigenstate properties}
The NV spin ground state Hamiltonian $H_{gs}$ can be written as:
\begin{gather} 
H_{gs} = H_e + H_n \nonumber \\
H_e = D_{gs}S_z^2 + \gamma_B (B_x S_x + B_z S_z)  \nonumber \\
H_n = \vec{I}\cdot\vec{A}\cdot\vec{S} + \gamma_N (B_x I_x + B_z I_z),  \nonumber
\end{gather} 
where $H_e$ and $H_n$ denote the Hamiltonians associated with the electron spin (S=1) and $^{15}$N nuclear spin ($I=\frac{1}{2}$),  respectively. $D_{gs}$ $\approx$ 2.87 GHz is the zero-field splitting,  $\gamma_B$ $\approx$ 2.87 MHz/G is the electron spin gyromagnetic ratio,  $\gamma_N$ $\approx$ 0.4316 kHz/G is the $^{15}$N nuclear spin gyromagnetic ratio, and $\vec{A}$ is the hyperfine tensor with only diagonal elements: $A_{xx} = A_{yy}$ $\approx$ 3.65 MHz and $A_{zz}$ $\approx$ 3.03 MHz \cite{Gali2009, felton2009hyperfine, smeltzer2009robust}. $S_{x,y,z}$ and $I_{x,y,z}$ are the spin-1 and spin-$\frac{1}{2}$ Pauli matrices, respectively. We applied the bias field $|B|>60$ G, such that $H_e$ always dominates over $H_n$ (at any $\theta_B$). The electron eigenstates, denoted by $|0, \pm\rangle$ throughout the paper, are thus mainly determined by $H_e$, and $H_n$ splits each state into two nuclear spin sublevels (see Supplementary Information). 

Under a bias magnetic field along $\hat{z}$ ($\theta_B = 0$),  i.e. the conventional magnetometry condition, the electron eigenstates are $|m_S = 0, \pm1\rangle$ (Fig. 1c). As the bias field direction rotates ($\theta_B \neq 0$), the eigenbasis changes. At $\theta_B = 90\degree$ ($B_z = 0$, $B_x > 0$), the eigenstates are: $|0\rangle \approx |m_S = 0\rangle, |-\rangle = \frac{1}{\sqrt{2}}(|m_S = +1\rangle - |m_S = -1\rangle), |+\rangle \approx \frac{1} {\sqrt{2}}(|m_S = +1\rangle + |m_S = -1\rangle)$, and the states $|\pm\rangle$ split in energy by $\approx \frac{\gamma_B^2 B_x^2}{D_{gs}}$ (Fig. 1c). Under large $B_x$, $|m_S = 0\rangle$ and $\frac{1} {\sqrt{2}}(|m_S = +1\rangle + |m_S = -1\rangle)$ are slightly hybridized in composing $|0\rangle$ and $|+\rangle$, hence the approximate equality in the above expressions. This hybridization results in finite $\langle S_x \rangle$ values for $|0\rangle$ and $|+\rangle$: $\langle S_x \rangle_{0 } < 0, \langle S_x \rangle_{+} > 0$ (see Supplementary Information).

\begin{figure}[t]
\centering
\includegraphics[scale=0.56]{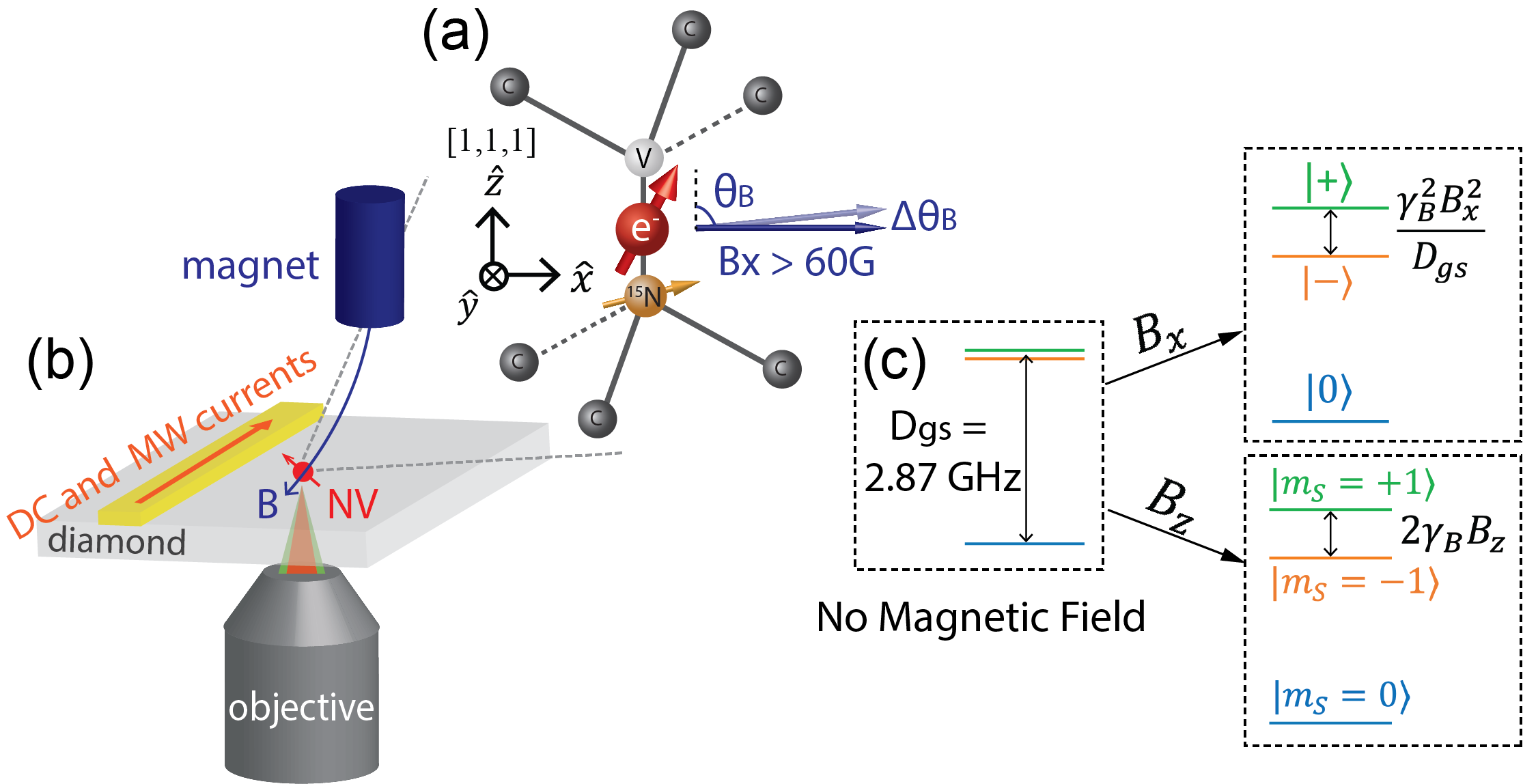}
\caption{\label{fig:epsart} Schematic of the NV center and experimental setup. (a) $^{15}$NV in the presence of a bias magnetic field acting perpendicular to the NV axis. The coordinate system is depicted on the left. The bias field direction rotates in the XZ plane.  $\theta_B$ is varied around $90\degree$ and $\Delta\theta_B$ is the signal to be detected. (b) A permanent magnet and DC currents in the metal stripline together control the magnetic field at the NV. (c) NV electron energy eigenstates under parallel ($B_z$) and perpendicular ($B_x$) magnetic fields.} 
\end{figure}

The states $|\pm\rangle$ are equal superpositions of $|m_S=\pm 1\rangle$ at $\theta_B = 90\degree$, hence $\langle S_z \rangle_{\pm} = 0.$ However, as $\theta_B$ deviates from $90\degree$, the $|m_S=\pm 1\rangle$ amplitudes are no longer equal. Due to the large zero-field splitting, the imbalance grows rapidly with the off-angle $\Delta\theta_B \equiv\theta_B-90\degree$, and consequently, $\langle S_z \rangle_{\pm}$ acquire finite values (Fig. 2a). On the other hand, $\langle S_x \rangle$ barely changes (Fig. 2b). The change in $\langle S_z \rangle$ dramatically affects the electron spin interaction with the nuclear spin, thus providing a unique way to sense $\Delta\theta_B$. 

\begin{figure}[!]
\centering
 \includegraphics[scale=0.52]{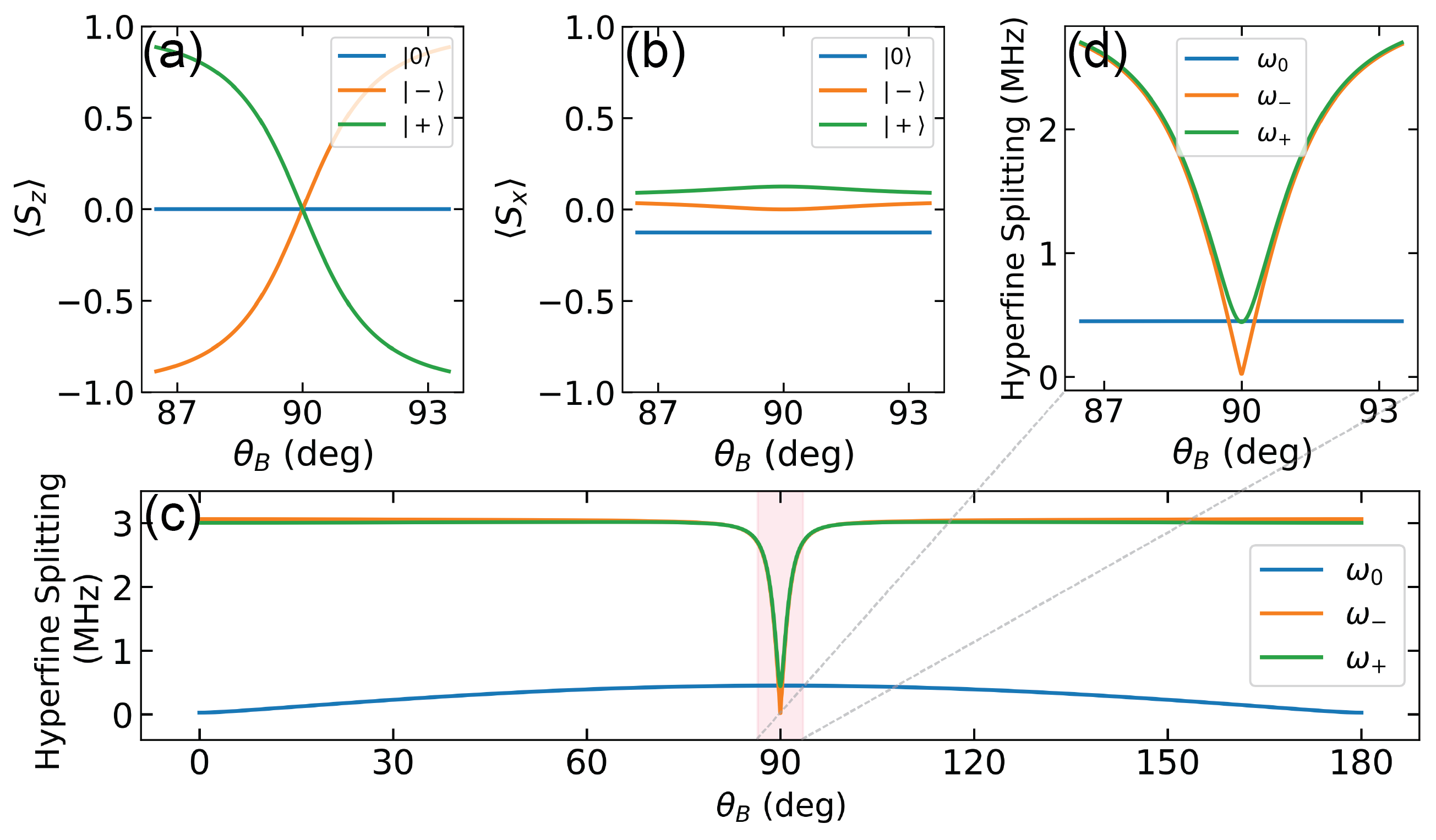}
\caption{\label{fig:epsart} 
(a)(b) Angle dependence of the electron spin operator expectation values $\langle\vec{S}\rangle$ at each state $|0,\pm\rangle$, calculated under $|B|$ = 65 G. $\langle S_y\rangle=0$ for all states, thus is not plotted. (c)(d) Angle dependence of the nuclear spin sublevel splittings $\omega$ at each electron state. (c) zooms in on the red shaded area around $90\degree$ in (d). $\omega_-$, the orange curve in (d), sharply increases with the off-angle $\Delta \theta_B$, serving as the key to our angle sensing approach. 
}
\end{figure}

\subsection*{Angle-dependent hyperfine interaction} 
A given electron state exerts effective hyperfine fields at the nuclear spin, determined by its spin operator expectation values. Specifically, the nuclear spin Hamiltonian is $H_n = A_{||} I_x \langle S_x \rangle_{|\psi_e\rangle} + A_{\perp} I_z  \langle S_z \rangle_{|\psi_e\rangle} + \gamma_N \left(B_x I_x + B_z I_z \right)$
, where $|\psi_e\rangle = |0,\pm\rangle$. $H_n$ splits each electron state ($|0,\pm\rangle$) into two nuclear sublevels and the splitting energy ($\hbar \omega$) can be obtained by diagonalizing $H_n$. Fig. 2c plots $\omega$ as a function of $\theta_B$ calculated under $|B|=65$ G and Fig. 2d zooms in on a small angle range centered at $90\degree$. The $|-\rangle$ state splitting ($\omega_-$) is especially interesting: it grows linearly with $\Delta\theta_B$. The slope $\frac{d\omega_-}{d\theta_B}$, as we will see soon, directly determines the angle sensitivity. 

\subsection*{Nuclear spin assisted angle-sensing} 
We now demonstrate detection of small angle changes using the angle-sensitive hyperfine interaction. Either $\omega_-$ or $\omega_+$ can be used, as they both change with $\theta_B$. We choose to use $\omega_-$ since its angle dependence is steeper. To measure this quantity, we performed electron spin-echo interferometry, where the spin-echo signal is dramatically affected by the hyperfine splitting due to the electron-spin-echo-envelope-modulation (ESEEM) effect \cite{Rowan1965, Childress2006, Maze2008, Ohno2012}.

A typical spin-echo sequence is shown in Supplementary Information Fig. 3. The electron is first prepared in a superposition of $|0\rangle$ and $|-\rangle$, and then accumulates phase during the free evolution time $\tau$ between the two $\pi/2$ pulses. The $\pi$ refocusing pulse decouples the fields at frequencies other than $1/\tau$.

The ESEEM effect occurs when the nuclear spin undergoes Larmor precession, with the frequency ($\omega_0$ or $\omega_-$) conditioned on the electron state ($|0\rangle$ or $|-\rangle$). Consequently, the electron and nuclear spins are periodically entangled and disentangled at a rate determined by $\omega_0$ and $\omega_-$. The spin-echo amplitude measures the electron coherence, which is directly affected by its entanglement with the nuclear spin, hence exhibiting collapses and revivals. As $\omega_-$ is highly sensitive to $\theta_B$, the spin-echo signal shows angle-dependent modulation patterns. A more detailed analysis is given in Supplementary Information, where the spin-echo signal $P$ obtains a simple expression: $P = 1-|\hat{\omega}_0 \times \hat{\omega}_-|^2  \sin^2 \left( {\frac{\omega_0 \tau}{4}} \right)  \sin^2 \left({\frac{\omega_- \tau}{4}} \right)$.

\begin{figure}[!]
\includegraphics[scale=0.645]{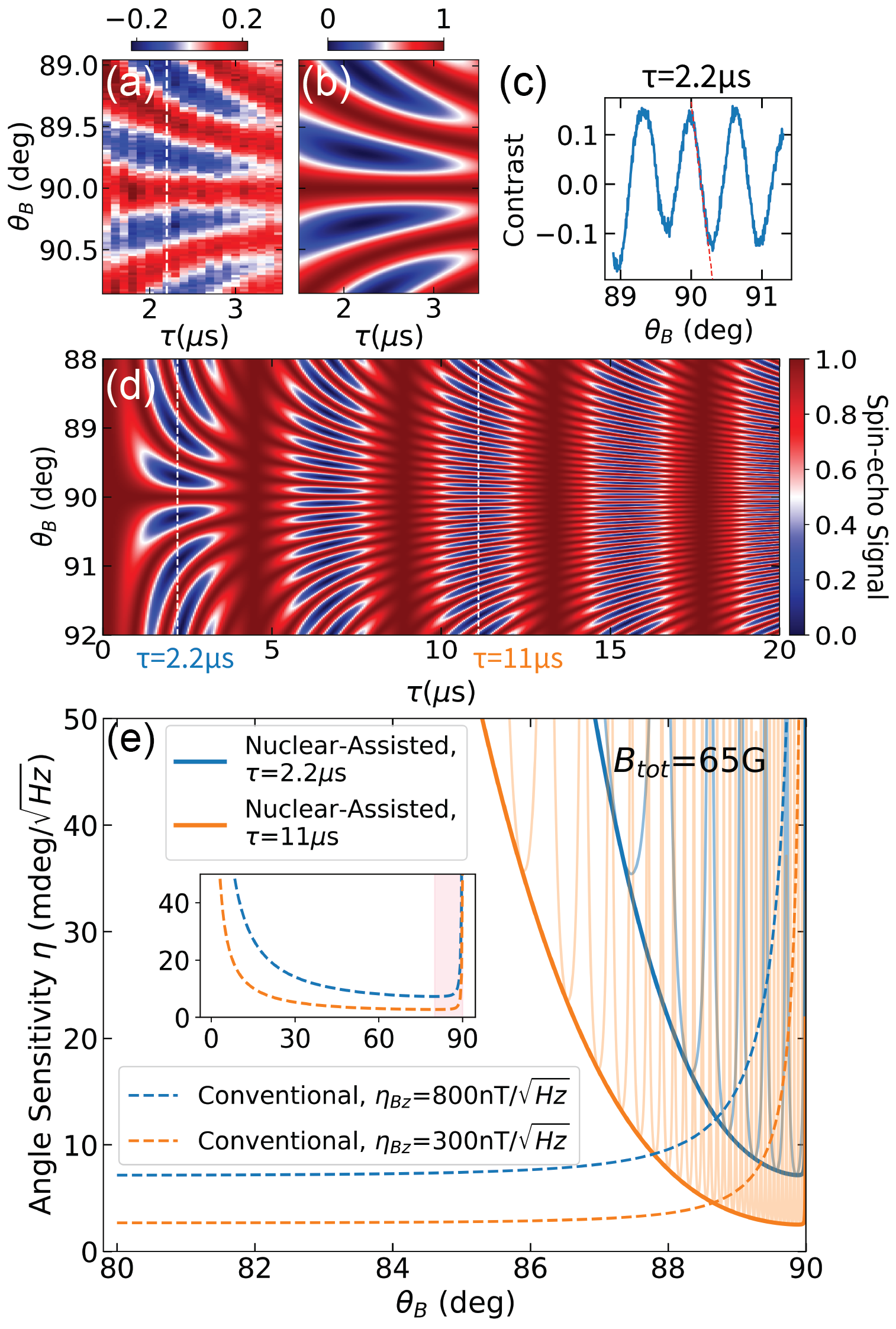} 
\caption{\label{}Demonstration of nuclear spin assisted angle sensing using spin-echo interferometry. (a) Experimental and (b) theoretical spin-echo signal as a function of the free evolution time $\tau$ and $\theta_B$ under $|B|\approx65$ G. The colorbar represents the fluorescence contrast in (a) and the spin-echo amplitude in (b). (c) The experimental spin-echo signal at fixed $\tau = 2.2$ $\mu s$ (white dashed line in (a)) and the red dashed line denotes the largest slope as $\theta_B$ varies.  (d) Spin-echo calculation in broader $\theta_B$ and $\tau$ ranges, where the two white dashed lines mark the fixed $\tau$ times at which sensitivities are plotted in (e). (e) Comparison of angle sensitivities using the nuclear-assisted method (solid curves) in this work vs. conventional magnetometry (dashed curves) based on the electron Zeeman interaction. The nuclear-assisted sensitivities $\eta$ oscillate with $\theta_B$, as shown by the semi-transparent solid curves, while the thick curves sketch the envelopes $\eta^*$. They are compared with conventional approaches assuming two different $B_z$ sensitivities ($300 \frac{nT}{\sqrt{Hz}}$ and $800 \frac{nT}{\sqrt{Hz}}$) under a parallel bias field. Inset: the conventional magnetometry sensitivity $\eta_{con}$ in the full angle range.}
\end{figure}

We performed spin-echo experiments between $|0\rangle$ and $|-\rangle$ under $|B|\approx65$ G, as $\theta_B$ varied between $89\degree$ and $91\degree$ (Fig. 3a). 
It shows good agreement with the above expression for $P$ (Fig. 3b). At fixed $\tau=2.2$ $\mu s$, the spin-echo signal can sensitively detect small angle changes at the largest slope (red dashed line in Fig. 3c). The corresponding sensitivity is $\sim 13$ $\frac{mdeg}{\sqrt{Hz}}$, provided the single NV fluorescence $\sim100$ kcps and optical contrast $\sim 15\%$ in the experiment (see Supplementary Information). To get an overall picture, Fig. 3d expands on Fig. 3b, showing the spin-echo signal $P$ in broader $\theta_B$ and $\tau$ ranges.

Taking the derivative of $P$ with respect to $\theta_B$, we obtain an analytical expression of the angle sensitivity $\eta$ (Supplementary Information):
\begin{gather}
\eta =  \frac{\eta^*}{\vert\sin^2 \left( {\frac{\omega_0 \tau}{4}} \right) \cdot \sin \left({\frac{\omega_- \tau}{2}} \right)\vert} \\
\eta^* \equiv \frac{4}{\gamma_{\theta} C}  \sqrt{\frac{1}{FT_r \tau} \cdot \frac{t_{ini} + \tau}{\tau}}, 
\end{gather}
where $F$ represents the NV fluorescence, $C$ the optical contrast of different spin states, $t_{ini}$ the spin initialization time and $T_r$ the spin state readout time. $\gamma_{\theta}\equiv \frac{d\omega_-}{d\theta_B}$ denotes the slope of the angle-dependent hyperfine interaction (Fig. 2d). Note that $\gamma_{\theta}$ is playing an analogous role as the gyromagnetic ratio in conventional magnetometry. The denominator in Eq. (1) causes modulation in $\tau$, i.e. the angle sensitivity is periodically lost and regained at different $\tau$ times (Fig. 3d), and $\eta^*$ is the modulation envelope. To optimize the sensitivity, we need to pick $\tau$ that maximizes $\vert\sin^2 \left( {\frac{\omega_0 \tau}{4}} \right) \cdot \sin \left({\frac{\omega_- \tau}{2}} \right)\vert$. As examples, sensitivities were evaluated at $\tau = 2.2$ $\mu s$ and $11$ $\mu s$ and plotted in Fig. 3e, represented by the blue and orange solid curves respectively.

On the other hand, conventional magnetometry also detects magnetic field angle changes via the electron Zeeman interaction. Its angle sensitivity is proportional to the static magnetic field sensitivity along $\hat{z}$: $\eta_{con} = \frac{\eta_{Bz}}{|B| \sin \theta_B}$. With a parallel bias magnetic field, $\eta_{Bz}$ is typically between tens of $\frac{nT}{\sqrt{Hz}}$ and a few $\frac{\mu T}{\sqrt{Hz}}$ depending on experimental parameters \cite{schirhagl2014nitrogen, casola2018probing}, and $\eta_{Bz}$ decreases as the bias field turns toward a perpendicular direction (see Supplementary Information). Fig. 3e plots $\eta_{con}$ as a function of $\theta_B$ assuming $\eta_{Bz} $ is originally $300$ or $800 \frac{nT}{\sqrt{Hz}}$ under a parallel bias field.

As illustrated in Fig. 3e, our nuclear-assisted approach and conventional magnetometry work in complimentary regimes. The conventional method works well until $\theta_B$ approaches $90\degree$ (Fig. 3e inset), when $\frac{\gamma_B^2 B_x^2}{D_{gs}} \gtrsim 2\gamma_BB_z$ so the electron eigenbasis changes from $|m_S = 0,\pm 1\rangle$ to $|0, \pm \rangle$, and after that the nuclear-assisted approach takes over. The sensitivities of both methods are limited by low-frequency noise, up to different effective coupling constants (see Supplementary Information). 

\subsection*{Detection of anisotropic noise} 
The NV couples to magnetic field noise through its electron spin operators. Similar to the hyperfine interaction, the noise coupling strength, and hence the spin coherence, also exhibits strong angle dependence. We show that this provides a useful way to distinguish and characterize anisotropic noise. 
 
When the NV is in a superposition of $|0\rangle$ and $|\pm \rangle$, magnetic field noise $\delta \vec{B} (t)$ induces fluctuations of the transition energy: $\delta E_{\pm, 0}(t) = \delta \vec{B} (t)\cdot (\langle \vec{S} \rangle_{\pm} - \langle \vec{S} \rangle_{0 })$.
It can be shown that the coherence is affected by the variance of the fluctuation $\langle \delta E^2 \rangle$ (see Supplementary Information). Recalling $\langle S_z\rangle$ ($\langle S_x \rangle$) in Fig. 2a (b), we get:
\begin{gather}
\langle \delta E_{-0}^2 \rangle = \langle \delta B_x^2 \rangle\langle S_x\rangle_{0}^2  +  \langle \delta B_z^2 \rangle\langle S_z\rangle_{-}^2 - 2 \langle \delta B_x \delta B_z \rangle \langle S_x\rangle_{0} \langle S_z\rangle_{-}  \\
\langle \delta E_{+0}^2 \rangle = 4 \langle \delta B_x^2 \rangle \langle S_x\rangle_{0}^2 +  \langle \delta B_z^2 \rangle \langle S_z\rangle_{+}^2-4 \langle \delta B_x \delta B_z \rangle \langle S_x\rangle_{0} \langle S_z\rangle_{+} 
\end{gather}

The last terms in Eqs. (3, 4) suggest the coherence is sensitive to the correlation between $\delta B_x$ and $\delta B_z$, which is non-zero for anisotropic noise. Since $\langle S_z\rangle_{\pm} $ is an odd function of $\Delta \theta_B$, while $\langle S_x\rangle_{0 }$ an even function, depending on the sign of $\langle \delta B_x \delta B_z \rangle$, the coherence is longer on one side than the other around $\theta_B=90\degree$, and the states $|\pm\rangle$ show opposite asymmetry. If the noise is isotropic, i.e. $\langle \delta B_x \delta B_z \rangle = 0$, the coherence is then symmetric around $90\degree$. Therefore, by examining the angle dependence of the coherence time, we can distinguish the anisotropic noise and further characterize its direction based on the asymmetry. This directional information, on the other hand, cannot be obtained under a parallel bias magnetic field ($\theta_B \approx 0\degree$), where the NV only couples to the $\hat{z}$ component of the noise.
  
\begin{figure}[t]
\includegraphics[scale=0.42]{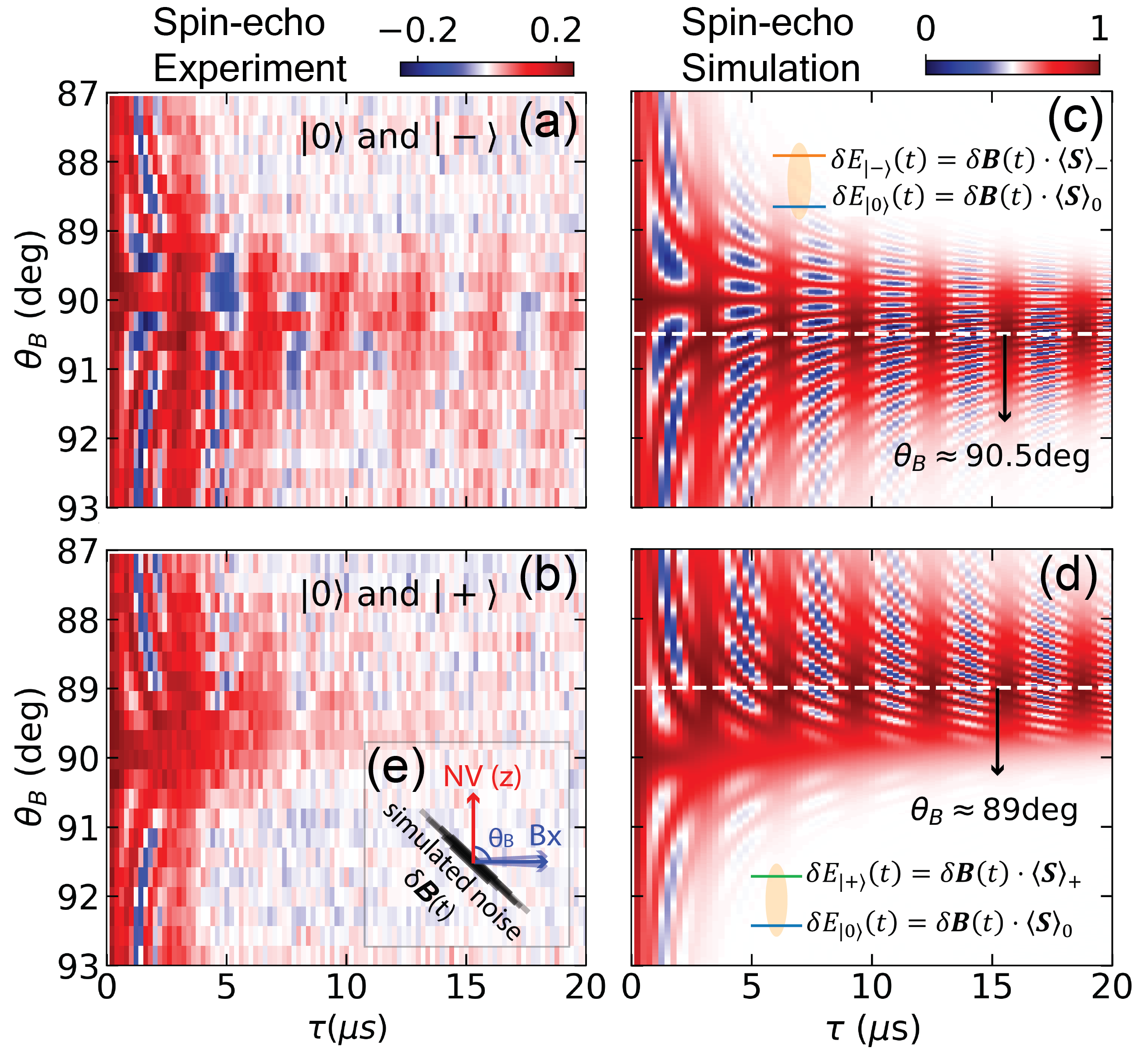} 
\caption{\label{} The NV coherence time exhibits asymmetry between $\theta_B < 90\degree$ and $\theta_B > 90\degree$, originating from anisotropic magnetic field noise. (a)(b) Spin-echo experiments between $|0\rangle$ and $|\mp\rangle$ as $\theta_B$ varies under $|B|\approx93$ G. Colorbar represents the fluorescence contrast. (c)(d) Simulated spin-echo signals with a magnetic field noise acting along a direction at $-45\degree/+135\degree$ from the $+\hat{x}$ axis, as depicted in (e). The white dashed lines mark the critical angles at which the noise coupling is maximally suppressed. Colorbar represents the spin-echo amplitude.}
\end{figure}
 
The coherence asymmetry observed in our experiment (Fig. 4a, b) indicates that the noise coupled to the NV is anisotropic. We argue (see Supplementary Information) that this is likely due to the dipolar interaction with a few nearby randomly-flipping spins (such as P1 centers or $^{13}$C nuclear spins). To further illustrate this effect, we performed spin-echo simulation under a fully anisotropic noise, where the noise only fluctuates along a straight line at $-45\degree/+135\degree$ relative to $+\hat{x}$ (Fig. 4e), such that $\delta B_x(t) + \delta B_z(t) = 0$. Under this condition, there exist critical angles at which the noise coupling is maximally suppressed (white dashed lines in Fig. 4c, d).

\section*{Discussion}
While conventional NV magnetometry fails when the bias magnetic field orients perpendicular to the NV axis, here we demonstrated a new strategy which uses the electron eigenbasis change for sensing based on the hyperfine interaction with the nuclear spin. The achieved sensitivity ($\eta \approx 13$ $\frac{mdeg}{\sqrt{Hz}}$) can be further improved by using cleaner diamond samples with less noise (e.g. by using $^{12}$C enriched diamonds \cite{balasubramanian2009ultralong} or chemical termination to reduce surface spins \cite{sangtawesin2019origins}), increasing photon collection efficiency (e.g. by fabricating microlens \cite{Siyushev2010, Hadden2010, Castelletto2011, grote2017imaging, Robledo2011} or pillars \cite{babinec2010diamond, hausmann2010fabrication, Maletinsky2012, appel2016fabrication, Zhou2017}) or using NV ensembles \cite{bauch2018ultralong, barry2020sensitivity}. For example, as the sensitivity is improved by the square root of the number of NVs, a typical ensemble density of $10^{12}/cm^2$ can achieve $\eta < 0.3$ $\frac{mdeg}{\sqrt{Hz}}$. Our work expands the already remarkable sensing versatility of the NV center. Note that the NV is also a sensitive electrometer under a perpendicular magnetic field \cite{Dolde2011, Michl2019, iwasaki2017direct, mittiga2018imaging}. In addition, this general idea of sensing by measuring the interaction with an ancillary sensor may be extended to other quantum sensing platforms. 

We also demonstrated a method to distinguish and characterize anisotropic noise in the XZ plane (i.e. $\delta B_x$ and $\delta B_z$) based on the asymmetric angle dependence of coherence time. By further rotating the bias field in the XY plane (perpendicular to the NV axis), we could obtain information along all directions, which can help locate noise sources at the nanoscale. Moreover, we showed that the anisotropic component of the noise can be maximally decoupled by applying a bias field at the critical angle, which may allow a great improvement of the NV spin coherence.

\section*{Methods}
The diamond sample was an electronic-grade CVD diamond (Element Six), with a natural abundance (1.1$\%$) of $^{13}$C impurity spins. Negatively-charged NV centers were created by ion implantation (18 keV) followed by vacuum annealing at 800 $\degree$C for two hours. NV experiments were performed on a home-built confocal laser scanning microscope. A 532 nm green laser was used to initialize the NV spin to the $m_S = 0$ state and generate spin-dependent photoluminescence for optical readout.  

The spin-echo simulation was done using the QuTip (Quantum Toolbox in Python) software \cite{Johansson2012, Johansson2013}. To simulate the coupling with the anisotropic magnetic noise, collapse operators were defined, and the time evolution was governed by the Lindblad master equation.  

\nocite{*}
\bibliography{manuscript_reference}

\section*{Data Availability}
The data generated and analyzed during this study are available from the authors upon reasonable request.

\section*{Code Availability}
The code used for simulation presented in this study is available from the authors upon reasonable request.

\section*{Acknowledgments}
We gratefully thank Norman Y. Yao for valuable discussions and advice. Z. Q. also thanks Matthew Turner for diamond annealing assistance. This work was primarily supported by ARO Grant No. W911NF-17-1-0023. Fabrication of samples was supported by the U.S. Department of Energy, Basic Energy Sciences Office, Division of Materials Sciences and Engineering under award DE-SC0019300. A.Y. also acknowledges support from the STC Center for Integrated Quantum Materials, NSF Grant No. DMR-1231319, and the Aspen Center of Physics supported by NSF grant PHY-1607611.  

\section*{Contributions}
A. Y. and Z. Q. conceived the project.  Z. Q. performed the experiments, analyzed data, and produced figures. Z. Q., U. V., A. H. and A. Y. discussed the results and contributed to the writing of the manuscript. A. Y. supervised the project. 

\end{document}


\title{Supplementary Information for \\ Nuclear Spin Assisted Magnetic Field Angle Sensing}

\author{Ziwei Qiu}
\affiliation{Department of Physics, Harvard University, Cambridge, MA 02138, USA}
\affiliation{John A. Paulson School of Engineering and Applied Sciences, Harvard University, Cambridge, MA 02138, USA}
\author{Uri Vool}
\affiliation{Department of Physics, Harvard University, Cambridge, MA 02138, USA}
\affiliation{John Harvard Distinguished Science Fellows Program, Harvard University, Cambridge MA 02138, USA}
\author{Assaf Hamo}
\affiliation{Department of Physics, Harvard University, Cambridge, MA 02138, USA}
\author{Amir Yacoby}
\email[Correspondence to: yacoby@physics.harvard.edu]{}
\affiliation{Department of Physics, Harvard University, Cambridge, MA 02138, USA}
\affiliation{John A. Paulson School of Engineering and Applied Sciences, Harvard University, Cambridge, MA 02138, USA}

\maketitle
\tableofcontents

\section{I. Experimental details}
\subsection{A. Diamond sample and NV characteristics}
Our experiment was performed on a type IIa diamond, with 1.1$\%$ naturally abundant $^{13}$C and of $2*2*0.05$ mm$^3$ in dimension, grown by chemical vapor deposition by Element Six. NV centers were created by $^{15}$N ion implantation (INNOViON) at 18 keV with a density of $30$ $/ \mu m^2$, and subsequent vacuum annealing for $\sim$2 hours at $800 \degree$C. The NV depth is estimated to be $\lesssim40$ nm below the surface. 

The NV used in the experiment was first characterized in the presence of a bias magnetic field oriented parallel to the NV axis, $B_{||} \approx $ 36 G. The measured dephasing time $T^*_{2, Ramsey} \lesssim 800$ ns and decoherence time $T_{2, Spin-echo}  \lesssim 1$ $\mu$s. The short coherence is attributed to various paramagnetic impurities in the environment, e.g. P1 center electron spins,  $^{13}$C nuclear spins and surface dangling bond spins. Multiple fabrication processes have been done on this sample before, which produce surface noise and degrade NV coherence. Employing dynamical decoupling techniques (such as Carr-Purcell-Meiboom-Gill sequences) with multiple equally spaced $\pi$ pulses can extend $T_2$ \cite{naydenov2011dynamical, de2010universal, ryan2010robust}. For example, with 128 $\pi$ pulses, $T_2$ can be extended to $\gtrsim 15$ $\mu$s, which indicates a finite correlation time of the noise.

The NV was then characterized under a perpendicular magnetic field, $B_{\perp} \approx93$ G. The original spin states $|m_S=0, \pm 1 \rangle$, are hybridized into new eigenstates $|0, \pm \rangle$, which are less sensitive to magnetic noise due to the reduced expectation values of electron spin operators $\langle S_{x, y, z} \rangle$. Consequently, the coherence time (between $|0\rangle$ and $|\pm\rangle$) increases. For the state $|-\rangle$, measured $T^*_{2, Ramsey, |-\rangle} \approx T_{2, Spin-echo, |-\rangle} \approx 5$  $\mu$s, and for $|+\rangle$, $T^*_{2,Ramsey, |+\rangle} \approx T_{2, Spin-echo, |+\rangle} \approx 2$  $\mu$s (see Fig. 4a-b of the main text). Note that the state $|-\rangle$ coherence time is longer. The reason is that $\langle S_x \rangle_{-} \approx 0$ while $\langle S_x \rangle_{+} > 0$ under a perpendicular magnetic field (Fig. 2b of the main text), hence $|-\rangle$ is less coupled to magnetic noise than $|+\rangle$. 

\subsection{B.  Experimental setup and control} 
A coplanar waveguide (CPW, made of Ti/Au 20 nm/200 nm) was fabricated directly on the diamond surface, patterned by electron-beam lithography. The central line is of $\sim 5$ $\mu$m in width and the NV is at a distance of $<10$ $\mu$m away from its edge. The diamond was placed on a glass coverslip and NVs were optically addressed with an NA=1.25 oil-immersion objective from beneath, as depicted in Fig. 1b of the main text. 532 nm laser of power $\sim 1.1$ mW, pulsed by an acoustic-optical modulator (AOM), was used to initialize and readout the NV spin states. The NV fluorescence was collected by an avalanche photodiode (APD). Pulsed microwave currents (generated from the source Agilent N9310A or Rhode\&Schwarz SMB100A) were delivered to the CPW to drive resonant transitions between different electron states. Pulse sequences were controlled by a TTL Pulse Generator (SpinCore PulseBlasterESR-PRO). 

A cylindrical NdFeB permanent magnet (6.35 mm in diameter and 12.7 mm in height), together with a small DC current ($<40$ mA) flowing in the central line (from a function generator Agilent 33120A), exerts a magnetic field at the site of NV. The magnet is mounted on stacked XYZ translation stages (Thorlabs TDC001 DC Servo Drive), which coarsely control the field magnitude and angle, and the DC current provides a fine angle control.

\section{II. Magnetic field angle sensing}
\subsection{A. NV electron energy eigenstates under a perpendicular magnetic field}

The NV electron spin Hamiltonian is dominated by the zero-field splitting and electron Zeeman interaction terms. Since the Hamiltonian is invariant under rotation about the NV symmetry axis, we define $\hat{z}$ to be along the NV axis and XZ the plane in which the external bias magnetic field lies.

When the bias field is exactly perpendicular to the NV axis, the NV spin ground state Hamiltonian is $H_{e} = D_{gs}S_z^2 + \gamma_B B_x S_x$ , where $D_{gs} \approx$ 2.87 GHz is the zero-field splitting (ZFS) due to spin-spin interaction and $\gamma_B \approx$ 2.8 MHz/G is the electron spin gyromagnetic ratio. Diagonalizing $H_{e}$ gives the eigenstates and eigenenergies (Fig. 1a).  As mentioned in the main text, the eigenstates, represented by $|0, \pm\rangle$, are superpositions of $|m_S = 0,\pm1\rangle$:
\begin{gather}
|0\rangle = |m_S = 0\rangle + \frac{\epsilon_1}{\sqrt{2}} \left({|m_S = +1\rangle + |m_S = -1\rangle}\right) \\
|-\rangle = \frac{|m_S = +1\rangle - |m_S = -1\rangle} {\sqrt{2}} \\
|+\rangle = \frac{1}{\sqrt{2}} \left({|m_S = +1\rangle + |m_S = -1\rangle}\right) + \epsilon_2  |m_S = 0\rangle
\end{gather}
Eq. (1) and (3) indicate that $|m_S = 0\rangle$ and $\frac{1}{\sqrt{2}} (|m_S = +1\rangle + |m_S = -1\rangle) $ are slightly hybridized in composing $|0\rangle$ and $|+\rangle$. As shown in Fig. 1b, $|\epsilon|\equiv|\epsilon_1| = |\epsilon_2|$ increases with $B_x$. Consequently, $\langle S_x\rangle_{0}\equiv\langle0|S_x|0\rangle < 0$, $\langle S_x\rangle_{-}\equiv\langle-|S_x|-\rangle > 0$, and their magnitudes increase with $B_x$ as well (Fig. 1c).

 \begin{figure}[h]
 \includegraphics[scale=0.4]{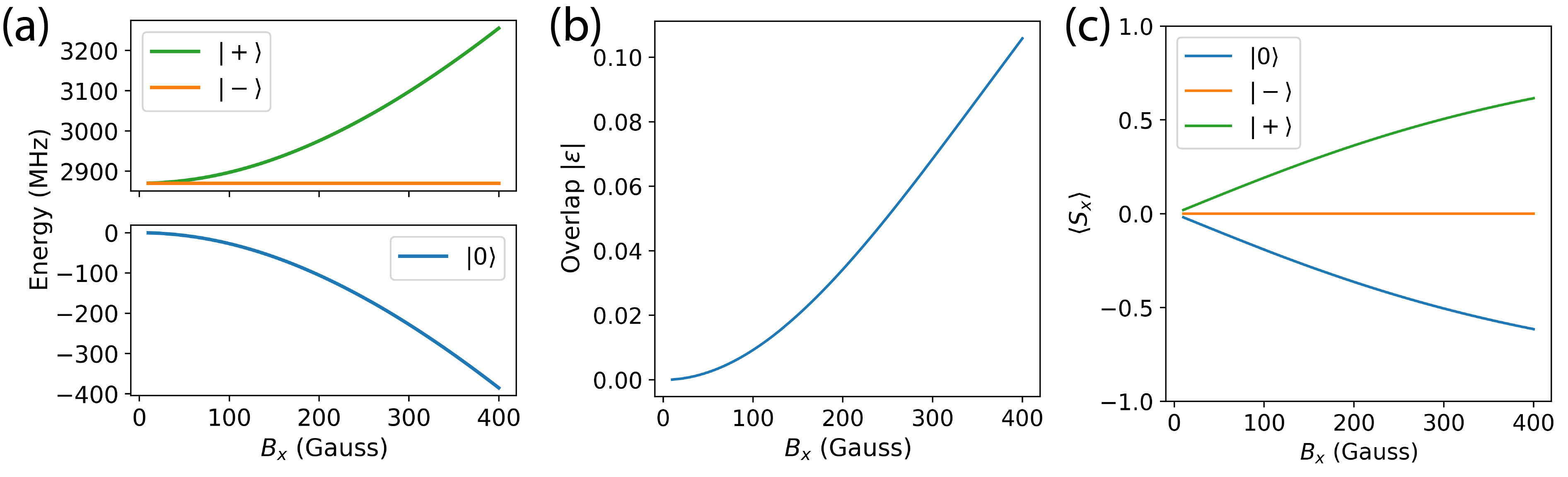}
 \caption{\label{} The electron eigenstates $|0, \pm\rangle$ under a bias magnetic field $B_x$ perpendicular to the NV axis. (a) Eigenstate energy levels as a function of the perpendicular field magnitude. The states $|\pm\rangle$ split in energy by $E_{\pm} \approx \gamma_B^2B_x^2/D_{gs}$,. (b) At large $B_x$, 
the states $|m_S = 0\rangle$ and $\frac{1}{\sqrt{2}} \left({|m_S = +1\rangle + |m_S = -1\rangle}\right)$ are hybridized in composing the eigenstates $|0\rangle$ and $|+\rangle$. The hybridization, quantified by $|\epsilon| \equiv |\epsilon_1| = |\epsilon_2|$ in Eq. (1) and (3) increases with $B_x$.  (c) $\langle S_x\rangle$ as a function of $B_x$.}
 \end{figure}

\subsection{B. Hyperfine interaction}

The hyperfine interaction with the $^{15}$N nuclear spin (I=1/2) splits each electron state $|0,\pm\rangle$ into two sublevels (Fig. 2a), and the splitting energy is sensitive to the magnetic field angle (Fig. 2c of the main text). Here we study the angle dependence of the nuclear spin quantization axis direction at each electron state, which affects the selection rules between different sublevels.

At a given electron state, the nuclear spin Hamiltonian is determined by the corresponding electron spin operator expectation values:
\begin{gather}
H_{n} =  \vec{I}\cdot\vec{A}\cdot\langle\vec{S}\rangle + \gamma_N \left(B_x I_x + B_z I_z \right) = A_{||} I_x \langle S_x \rangle + A_{\perp} I_z \langle S_z \rangle + \gamma_N \left(B_x I_x + B_z I_z \right),
\end{gather}
where $\langle S_y \rangle = 0$ is omitted, $\vec{A}$ is the hyperfine tensor and $\gamma_N \approx 0.4316$  kHz/G the $^{15}N$ is the nuclear spin gyromagnetic ratio. Diagonalizing $H_{n}$ gives the nuclear spin states and quantization axes ($\zeta$). In the following, $\theta_I$ denotes the angle between $\zeta$ and $\hat{x}$, defined as $\theta_I \equiv \arctan{\frac{\langle I_z \rangle}{\langle I_x \rangle}}$.

At the electron state $|0\rangle$, $\langle S_x \rangle_{0} < 0$ and $\langle S_z \rangle_{0} = 0$ (Fig. 2 of the main text), the first term $A_{||} I_x \langle S_x \rangle_{0}$ dominates. As the magnetic field angle ($\theta_B$) varies, $\langle S_x \rangle_{0}$ barely changes, therefore the splitting energy $\omega_0 \approx A_{||} |\langle S_x \rangle_0|$ is almost constant, and the quantization axis $\zeta_0$ always points along $\hat{x}$ (the blue line in Fig. 2b).

The electron state $|-\rangle$ is particularly interesting. $\langle S_x \rangle_{-} \approx 0$ regardless of angle (Fig. 2b of the main text), while $\langle S_z \rangle_{-}$ starts from 0 and grows rapidly with the off-angle $\Delta\theta_B$ (Fig. 2a of the main text). Consequently, the splitting energy $\omega_-$ increases with $\Delta\theta_B$,  which forms the basis of our angle sensing experiment presented in the main text, and the quantization axis $\zeta_-$ exhibits a very sharp angle dependence. At precisely $\theta_B=90\degree$,  i.e. $B_z = 0$ and $B_x\neq0$, the hyperfine interaction is almost zero so the nuclear Zeeman term becomes non-negligible. $\zeta_- $ points along the external bias field direction $\hat{x}$. However, as $\theta_B$ is slightly off from $90\degree$, $\langle S_z \rangle_{-}$ increases rapidly and hence $\zeta_- $ immediately turns toward $\hat{z}$ (the orange curve in Fig. 2b). The insets of Fig. 2b illustrate that a small magnetic field angle change results in a complete flip of the nuclear spin.

At the electron state $|+\rangle$,  $\langle S_x \rangle_{+}$ and $\langle S_z\rangle_{+}$ together determine the nuclear state. At $\theta_B=90\degree$, $\langle S_z \rangle_{+} = 0$, thus the splitting energy $\omega_+$ is determined by $\langle S_x \rangle_{+}$.  As $\theta_B$ is off from $90\degree$, $\langle S_z \rangle_{-}$ increases, giving rise to the angle dependence of $\omega_+$. The competition between $A_{||}I_x  \langle S_x \rangle_{+}$ and $A_{\perp} I_z  \langle S_z \rangle_{+}$ results in an angle-dependent quantization axis direction $\zeta_+$ (the green curve in Fig. 2b), but it is less sharp as compared to the state $|-\rangle$.

 \begin{figure}[h]
 \includegraphics[scale=0.35]{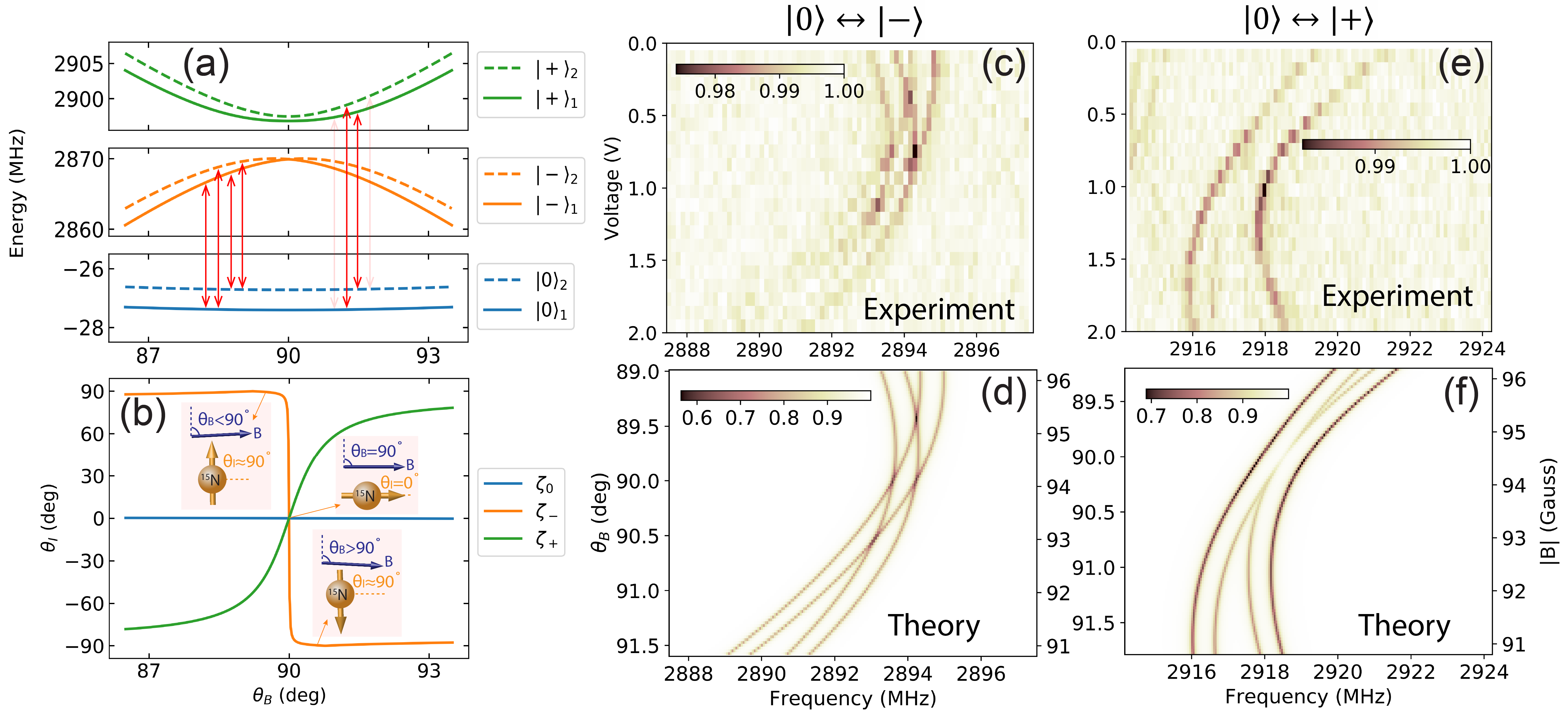}
 \caption{\label{} (a) Electron spin and nuclear spin energy levels as a function of the magnetic field angle $\theta_B$, calculated under $|B|=100$ G. The hyperfine splitting energy is sensitive to the angle. The red arrows denote all possible transitions between the electron states $|0\rangle$ and $|\pm\rangle$. The color brightness qualitatively indicates the transition efficiency under microwave driving. (b) The nuclear spin quantization axis direction $\zeta$ as a function of the bias magnetic field angle $\theta_B$ at each electron state (denoted by the subscripts in the legend). $\theta_I \equiv \arctan{\frac{\langle I_z \rangle}{\langle I_x \rangle}}$ is the angle between $\zeta$ and the $\hat{x}$ axis. Insets: illustration of the nuclear spin axis direction as $\theta_B$ varies in the proximity of $90\degree$ (c)(d) Experimental and theoretical electron spin resonance (ESR) spectra between $|0\rangle$ and $|-\rangle$ as $\theta_B$ varies. In the experiment, the magnetic field is controlled by modulating a DC current using voltage, and both the field angle and magnitude vary with the voltage. The ESR spectra are measured by Fourier transform of the Ramsey experiment. Colorbars represent the fluorescence contrast in the experiment, and arbitrary units in the calculation. (e)(f) The ESR spectra between $|0\rangle$ and $|+\rangle$. }  
 \end{figure}

Based on the above analysis, we can see that at a given magnetic field angle $\theta_B$, the nuclear spin quantization axis depends on the electron spin state, which affects the selection rules of the transitions between different electron spin states. For example, at $\theta_B$ slightly $<90\degree$, $\zeta_0$ points along $\hat{x}$ while $\zeta_-$ almost along $\hat{z}$, hence all the four sublevel transitions between $|0\rangle$ and $|-\rangle$ are allowed. More specifically, the driving efficiency of a transition is $\propto |\langle f |H'|i \rangle|^2$ ($H'$ is the coupling to the microwave fields, $|i\rangle$ and $|f\rangle$ are the initial and final states), as calculated and plotted in Fig. 3d-f as $\theta_B$ varies. The allowed transitions are experimentally measured by performing Fourier transform on Ramsey measurements between the electron states $|0\rangle$ and $|\pm\rangle$ (Fig. 3c-e). 

\subsection{C. Electron spin-echo envelope modulation (ESEEM)}

Our angle sensing approach presented in the main text is based on the angle-sensitive hyperfine interaction, measured by spin-echo sequences between $|0\rangle$ and $|-\rangle$ (Fig. 3). The essential point is that the electron spin-echo signal is dramatically affected by the hyperfine interaction due to the ESEEM effect, which is a well-studied phenomenon in NMR spectroscopy. 

 \begin{figure}[h]
 \includegraphics[scale=0.47]{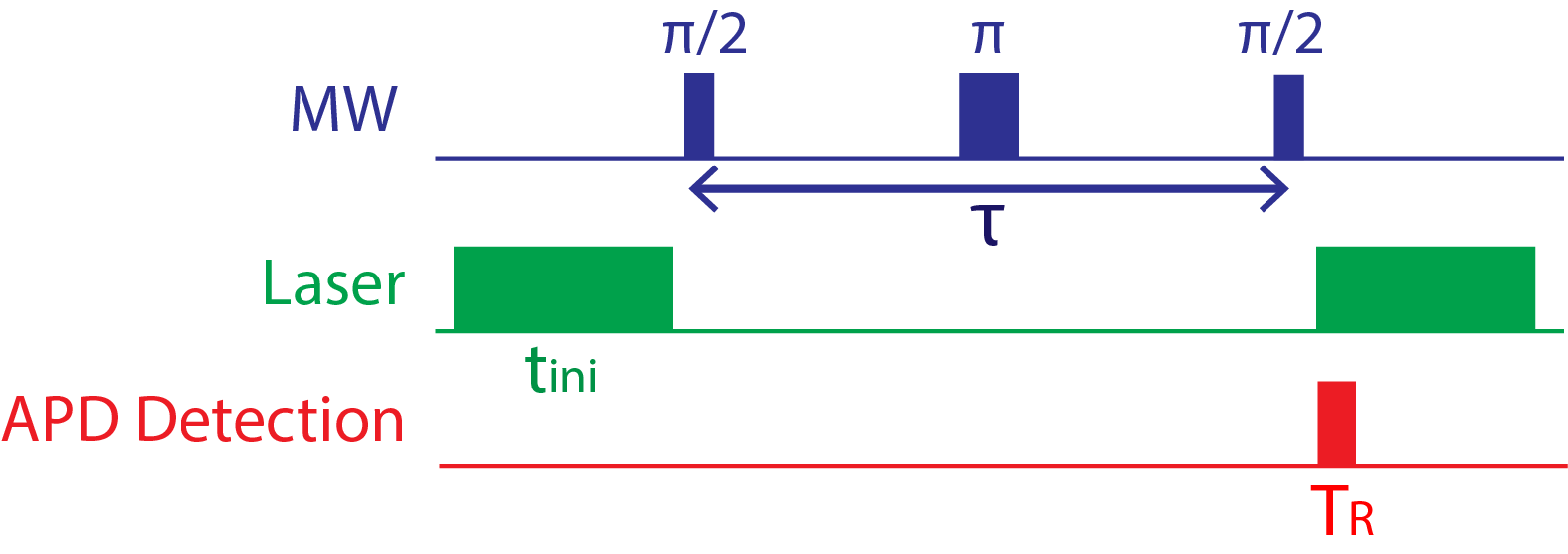}
 \caption{\label{} A typical spin-echo sequence. $t_{ini}$ is the laser pulse time for electron spin initialization, typically of a few $\mu$s. $T_r$ is the photon collection time for spin state readout, before the green laser reinitializes the spin state, typically of 200-300 ns. In our experiment, MW $\pi$ pulses are typically of $\sim 50$ns.}   
 \end{figure}

A detailed theoretical analysis of the ESEEM effect was first given by \cite{Rowan1965}, and recently people have studied it in NV centers due to various nuclear species \cite{Childress2006, Maze2008, Ohno2012}. Based on these former works, here presents a derivation aiming to obtain an analytical expression of the spin-echo signal.

Given the initial electron and nuclear spin states ($|\psi_{e}\rangle$ and $\rho_{N}$), the full density matrix is
\begin{eqnarray}
\rho_0= |\psi_{e}\rangle \langle \psi_{e}| \otimes \rho_{N}  = \frac{1}{2}(|0\rangle \langle 0| + |0\rangle \langle -| +& |-\rangle \langle 0|  + |-\rangle \langle -|)\otimes \rho_{N},
\end{eqnarray}
where $\rho_{N} =  \begin{bmatrix} 0.5 & 0 \\ 0 & 0.5 \end{bmatrix}$ represents the mixed nuclear spin state. The evolution operator $\hat{U_\tau}$ for a spin-echo sequence of free evolution time $\tau$ (Fig. 3) is:
\begin{eqnarray}
\hat{U}_\tau = \hat{U}(\tau/2)\:R_\pi\: \hat{U}(\tau/2) =e^{-iH\tau/2} \:e^{-i S_x \pi}\: e^{-iH\tau/2}, 
\end{eqnarray}
where $R_\pi$ represents the $\pi$ pulse and $\hat{U}(\tau/2)$ the free evolution operator for a duration of $\tau/2$. The electron spin-echo signal $P$ measures the overlap between the final and initial states:
 \begin{eqnarray}
P = \langle P_{\psi_0} \rangle = Tr(P_{\psi_0} \hat{U}_\tau \rho_0 \hat{U}_\tau^{\dagger}),
\end{eqnarray}
where $P_{\psi_0} = |\psi_{0}\rangle \langle \psi_{0}|$ is the projection operator. The Hamiltonian $H$ in the frame with respect to the electron spin states is $H = \vec{I} \cdot \vec{A} \cdot \vec{S} + \gamma_N \vec{B} \cdot \vec{I} $.
Plugging Eqs. (5), (6) into Eq. (7), we get:
\begin{eqnarray}
P = \frac{1}{2} \bigg\{ 1+Tr\left[\hat{U}_0\left(\frac{\tau}{2}\right) \hat{U}_-\left(\frac{\tau}{2}\right) \hat{U}_0^{\dagger}\left(\frac{\tau}{2}\right) \hat{U}_-^{\dagger}\left(\frac{\tau}{2}\right)\right] \bigg\}
\end{eqnarray}
$\hat{U}_0$ and $\hat{U}_-$ are the nuclear spin evolution operators corresponding to the electron spin states $|0\rangle$ and $|-\rangle$ respectively:
\begin{eqnarray}
\hat{U}_0 (t) \equiv e^{-iH_0t} = e^{-i (\hat{\omega}_0 \cdot \vec{I}) t} \\
\hat{U}_- (t) \equiv e^{-iH_-t} = e^{-i (\hat{\omega}_- \cdot \vec{I}) t},
\end{eqnarray}
where $\hat{\omega}_0$ and $\hat{\omega}_-$ are the unit vectors pointing along the corresponding nuclear spin quantization axes $\zeta_0$ and $\zeta_-$ (section II-B).
Plugging Eqs. (9), (10) into Eq. (8), we finally get:
\begin{eqnarray}
P = P(\theta_B, \tau) = 1-|\hat{\omega}_0 \times \hat{\omega}_-(\theta_B)|^2  \sin^2 \left( {\frac{|\hat{\omega}_0| \tau}{4}} \right)  \sin^2 \left({\frac{|\hat{\omega}_-(\theta_B)| \tau}{4}} \right),
\end{eqnarray}
where $\hat{\omega}_0$ and $\hat{\omega}_-$ are unit vectors along the quantization axes $\zeta_0$ and $\zeta_-$.

\subsection{D. Angle sensitivity analysis}
We detect small angle changes by measuring the spin-echo signal $P$ at a fixed $\tau$, hence the angle sensitivity $\eta$ is proportional to the derivative of $P$ with respect to $\theta_B$:
\begin{eqnarray}
\left(\frac{\partial P}{\partial \theta_B}\right)_{\tau} = \left( \frac{\partial P}{\partial |\omega_-|} \right)_{\tau} \left(\frac{\partial |\omega_-|}{\partial \theta_B}\right) = \gamma_{\theta} \left( \frac{\partial P}{\partial |\omega_-|} \right)_{\tau} 
\end{eqnarray}
The effective angle coupling constant $\gamma_{\theta} \equiv \frac{d \omega_-}{d \theta_B}$ corresponds to the slope of the hyperfine energy as a function of $\theta_B$ (Fig. 4).

\begin{figure}[h]
\includegraphics[scale=0.39]{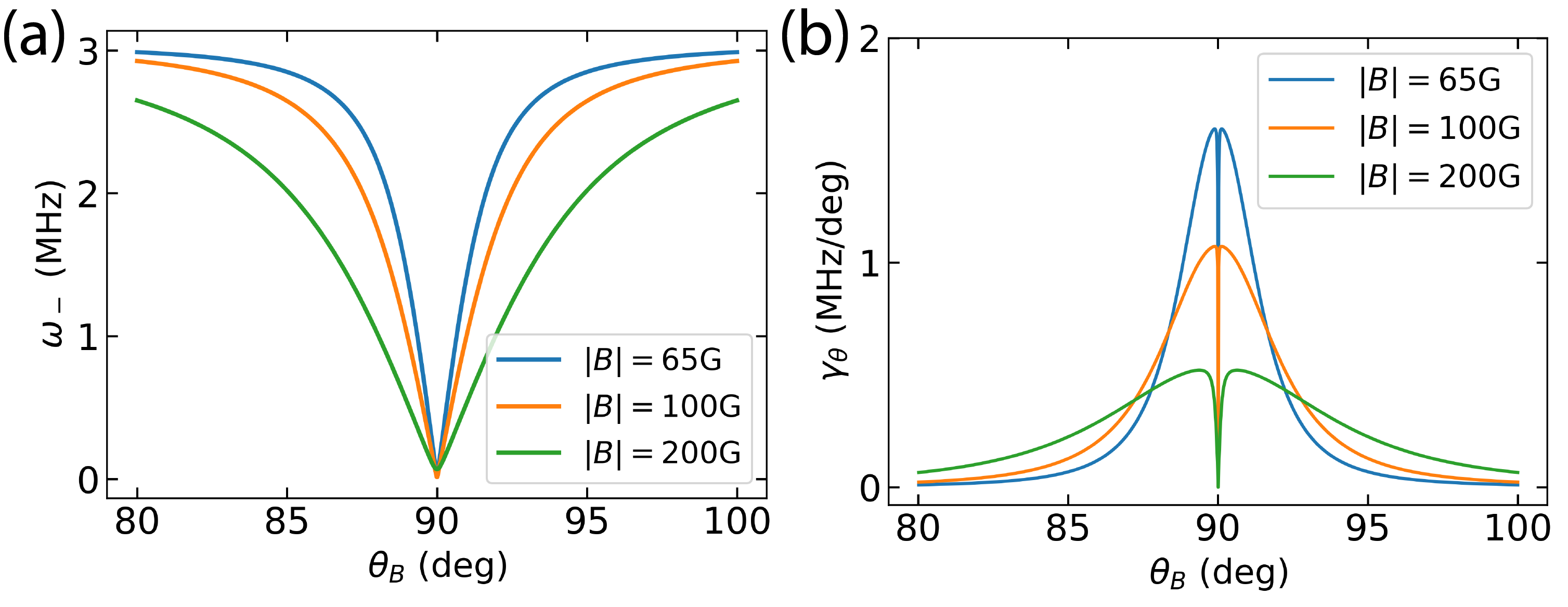}
\caption{\label{} (a) The $|-\rangle$ state hyperfine splitting $\omega_-$ as a function of $\theta_B$ under different magnetic field magnitudes. (b) The corresponding effective angle coupling constants $\gamma_{\theta} \equiv \frac{d \omega_-}{d \theta_B}$, which determine the angle sensitivity.} 
\end{figure}

In Eq. (11), $\hat{\omega}_0$ ($\zeta_0$) always points along $\hat{x}$ (the blue line in Fig. 3c). $\hat{\omega}_-$ ($\zeta_-$) is roughly along $\hat{z}$ but sharply turns toward $\hat{x}$ at $\theta_B=90\degree$ (the orange curve in Fig. 3c). This sharp transition, however, occurs in a very narrow angle range, exactly where $\frac{d \omega_-}{d \theta_B}$ almost diminishes at the minimum of $\omega_-$, thus it is not important to angle sensing. For the relevant angle range,  we can therefore take $|\hat{\omega}_0 \times \hat{\omega}_-|^2 \approx 1$. From Eq. (11), we have:
\begin{eqnarray}
\left( \frac{\partial P}{\partial |\omega_-|} \right)_{\tau} \approx -\frac{\tau}{4} \cdot \sin^2 \left( {\frac{\omega_0 \tau}{4}} \right) \cdot \sin \left({\frac{\omega_- \tau}{2}} \right)
\end{eqnarray}

Next, we discuss how to connect $\frac{\partial P}{\partial \theta_B}$ to the angle sensitivity $\eta$ by a proportionality constant determined by experimental parameters, including the NV fluorescence count rate $F$, the optical contrast $C$ between the states $|0\rangle$ and $|-\rangle$, the APD readout window $T_r$ and the initialization time $t_{ini}$ (Fig. 3). As our experiment was using a short $\tau$ ($\approx 2$ $\mu s$), comparable to the initialization time $t_{ini}$, we consider this measurement overhead. Suppose an angle signal $=\Delta\theta$, we expect to collect on average $\overline{N}_{photon}$ number of photons per spin-echo sequence:
\begin{eqnarray}
\overline{N}_{photon}=\biggr\vert\frac{\partial P}{\partial \theta_B}\biggr\vert \cdot \Delta\theta \cdot C \cdot FT_r
\end{eqnarray}
Each sequence is of length $= t_{ini} + \tau$, and it is repeated for $N_{avg} = \nicefrac{1}{(t_{ini} + \tau)}$ times within an integration time of 1 second. The expected total number of photons corresponding to the signal $\Delta\theta$ is:
\begin{eqnarray}
\textrm{signal} = \overline{N}_{photon}\cdot N_{avg} = \biggr\vert\frac{\partial P}{\partial \theta_B}\biggr\vert \cdot \Delta\theta \cdot C \cdot FT_r \cdot N_{avg}
 \end{eqnarray}
The photon shot noise is 
\begin{eqnarray}
\textrm{noise}=\sqrt{FT_r \cdot N_{avg}}
 \end{eqnarray}
Thus we obtain the signal-to-noise ratio (SNR) within 1 sec integration time:
\begin{eqnarray}
\textrm{SNR} = \biggr\vert\frac{\partial P}{\partial \theta_B}\biggr\vert \cdot \Delta\theta \cdot C \cdot \sqrt{FT_r \cdot N_{avg}} = \biggr\vert\frac{\partial P}{\partial \theta_B}\biggr\vert \cdot \Delta\theta \cdot C \cdot \sqrt{\frac{FT_r}{t_{ini} + \tau}}
 \end{eqnarray}
 The angle sensitivity $\eta$ is defined as the minimal angle $\Delta\theta$ that can be detected with SNR = 1 within 1 sec, hence
 \begin{eqnarray}
 \eta = \frac{1}{\vert\frac{\partial P}{\partial \theta_B}\vert  C} \sqrt{\frac{t_{ini} + \tau}{FT_r}}
 \end{eqnarray}
Putting together Eqs. (12) ,(13), (18), we get:
 \begin{gather}
 \eta = \frac{1}{\vert\sin^2 \left( {\frac{\omega_0 \tau}{4}} \right) \cdot \sin \left({\frac{\omega_- \tau}{2}} \right)\vert} \cdot \frac{4}{\gamma_{\theta} C} \cdot \sqrt{\frac{1}{FT_r \tau} \cdot \frac{t_{ini} + \tau}{\tau}} = \frac{\eta^*}{\vert\sin^2 \left( {\frac{\omega_0 \tau}{4}} \right) \cdot \sin \left({\frac{\omega_- \tau}{2}} \right)\vert} \\
 \eta^*\equiv \frac{4}{\gamma_{\theta} C} \cdot \sqrt{\frac{1}{FT_r \tau} \cdot \frac{t_{ini} + \tau}{\tau}}
 \end{gather}
As seen, $\eta$ is inversely proportional to the effective angle coupling constant $\gamma_{\theta}$. The term $\frac{1}{\vert\sin^2 \left( {\frac{\omega_0 \tau}{4}} \right) \cdot \sin \left({\frac{\omega_- \tau}{2}} \right)\vert}$ causes a fast modulation on the sensitivity, and $\eta^*$ is the modulation envelope. 
For example, supposing $F$ = 100 kcps, $C$ = 0.3, $T_r$ = 300 ns, $t_{ini}$ = 2 $\mu s$, which are typical conditions for bulk diamond single NV experiments, the sensitivities $\eta$ and $\eta^*$ as a function of $\tau$ are plotted in Fig. 5, under several different bias magnetic fields.

 \begin{figure}[h]
 \includegraphics[scale=0.4]{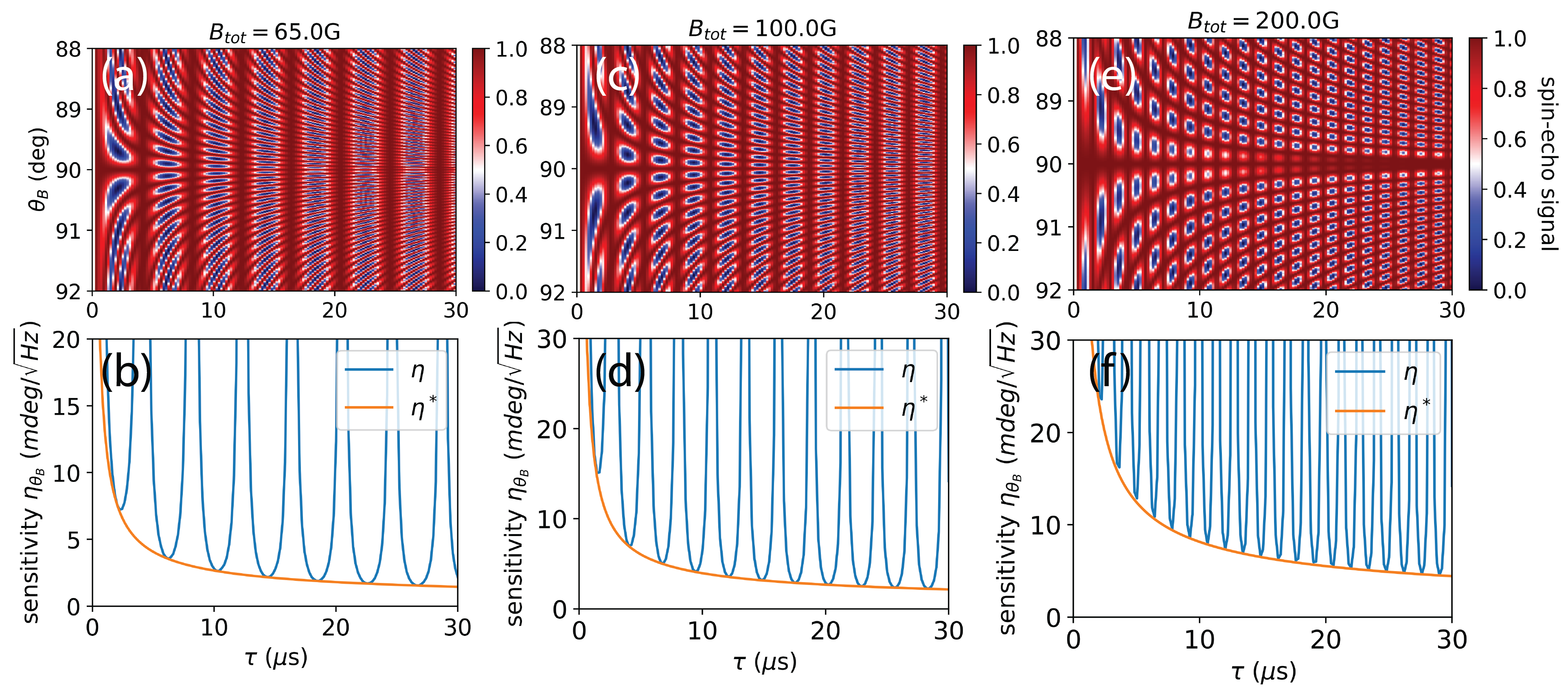}
\caption{\label{} Nuclear-assisted angle sensitivity under different magnetic fields $|B|$ = 65 G, 100 G and 200 G. (a)(c)(e) The calculated spin-echo signals as a function of $\tau$ and $\theta_B$, following Eq. (11). Colorbar represents the spin-echo amplitude. The derivative $\left(\frac{\partial P}{\partial \theta_B}\right)_{\tau}$ determines the angle sensitivity. (b)(d)(f) The corresponding angle sensitivities $\eta$ and envelopes $\eta^*$. Calculation uses the largest derivative at each $\tau$.} 
 \end{figure}

\subsection{E. Angle sensing with conventional NV magnetometry} 

In this section, we discuss using the conventional NV magnetometry based on the electron Zeeman interaction to detect magnetic field angle changes. 

When the external bias field is parallel to the NV axis ($\hat{z}$), the electron spin is in the basis of $|m_S = 0, \pm1\rangle$, hence the energy levels are sensitive to magnetic field signals along $\hat{z}$ ($\Delta B_z$) with the coupling constant $\gamma_{B_z} = \gamma_B$ (electron gyromagnetic ratio) $= 2.8$ MHz/G. The sensitivity $\eta_{B_z}$ is given by \cite{taylor2008high}:
\begin{eqnarray}
\eta_{B_z} \approx \frac{2}{\pi \gamma_{B_z} C} \cdot \sqrt{\frac{1}{FT_r \tau} \cdot \frac{t_{ini} + \tau}{\tau}}
\end{eqnarray}
As the bias field gradually turns toward a direction perpendicular to the NV axis, the electron basis changes and the energy levels become less sensitive to the signal $\Delta B_z$. That is, the $B_z$ coupling constant $\gamma_{Bz} \propto \langle S_z \rangle$ decreases as $\theta_B$ approaches $90\degree$ (Fig. 6 red curve). 

 \begin{figure}[h]
 \includegraphics[scale=0.45]{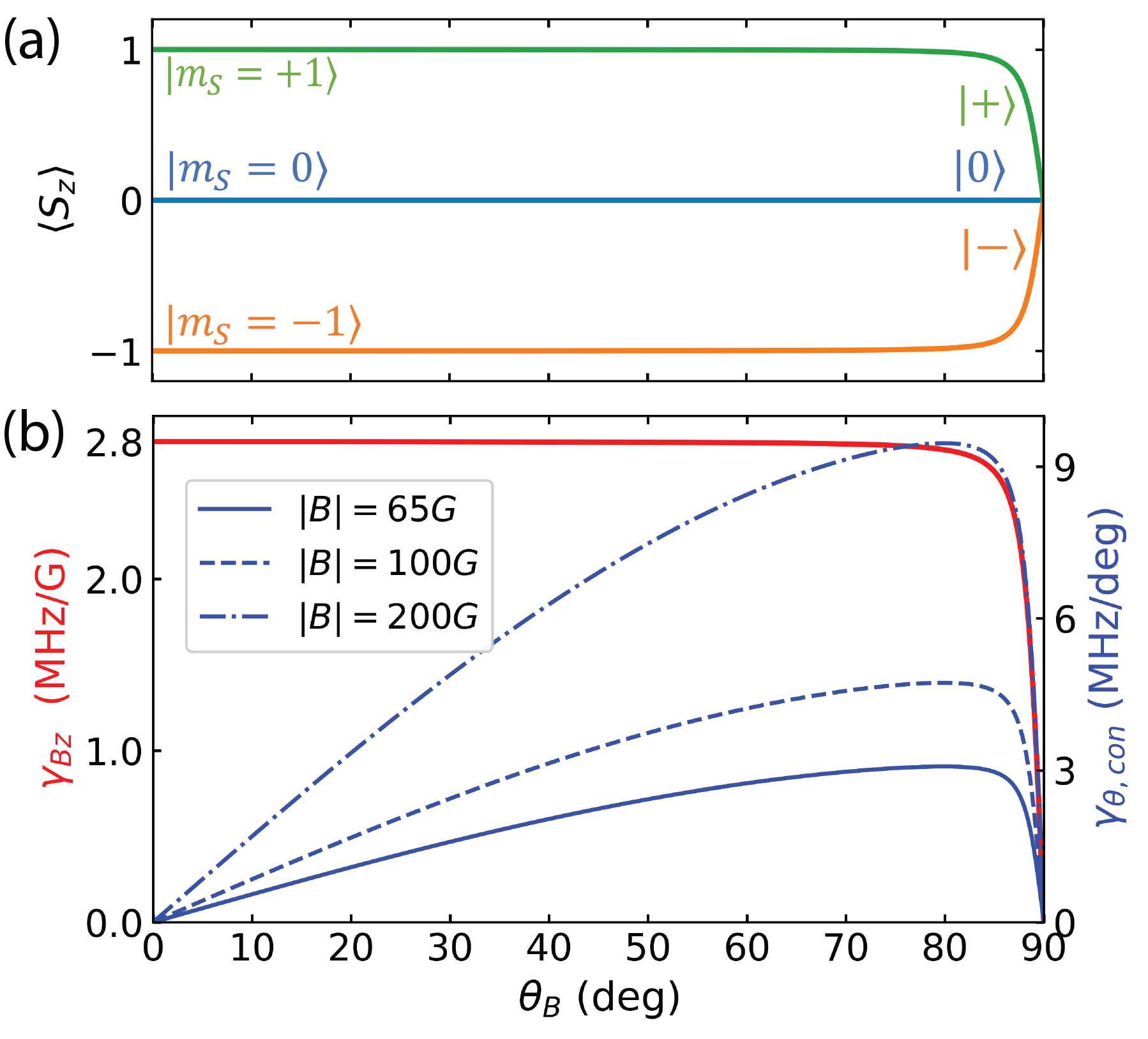}
\caption{\label{} (a) $\langle S_z \rangle$ diminishes as $\theta_B$ approaches $90\degree$. (b) The $B_z$ coupling constant $\gamma_{B_z}$ decreases with $\theta_B$ (red). The angle coupling constant $\gamma_{\theta, con.}$ first increases with $\theta_B$ but eventually decreases to zero (blue), under three different fixed bias magnetic field magnitudes $|B|$ (65, 100 and 200 G). } 
 \end{figure}

Since $B_z = |B| \cos{\theta_B}, \delta B_z = -|B| \sin{\theta_B} \cdot \delta \theta_B$, the angle sensitivity can be written as 
\begin{gather}
\eta_{\theta_B, con.} = \frac{\eta_{B_z}}{\sin{\theta_B} \cdot |B|} \approx \frac{2}{\pi \cdot \gamma_{B_z} \sin{\theta_B}  |B| \cdot C} \cdot \sqrt{\frac{1}{FT_r \tau} \cdot \frac{t_{ini} + \tau}{\tau}} \nonumber \\
\;\;\;\;\;\;\;\;\;\;\;\;\;\;\;\;\; = \frac{2}{\pi \gamma_{\theta, con.}  C}  \sqrt{\frac{1}{FT_r \tau} \cdot \frac{t_{ini} + \tau}{\tau}} \\
\gamma_{\theta, con.} \equiv  \gamma_{B_z} \sin{\theta_B}  |B| \;\;\;\;\;\;\;\;\;\;\;\;\;\;\;\;\;
\end{gather}
$\gamma_{\theta, con.}$ is the effective angle coupling constant. As $\theta_B$ approaches $90\degree$, $\gamma_{\theta, con.}$ initially increases due to $\sin{\theta_B}$, but eventually decreases to zero because $\gamma_{Bz} \propto \langle S_z \rangle$ diminishes (Fig. 6b blue curves). In terms of the sensitivity $\eta_{\theta_B, con.}$, as $\theta_B$ increases from 0 to $90\degree$, $\eta_{\theta_B, con.}$ first decreases until $\gamma_B^2B_x^2/D_{gs} \sim \gamma_B B_z$, at which the electron eigenbasis significantly changes, and after that $\eta_{\theta_B, con.}$ increases and the nuclear-assisted angle sensing approach takes over (see Fig. 3e of the main text). Comparing Eq. (20) and (21), we can see that the sensitivities of the two approaches take the same form, but differ by the effective angle coupling constant and proportionality factor. 

\section{III. Detection and origin of anisotropic noise} 
Here we discuss in more detail why the NV coherence time under a nearly perpendicular magnetic field is sensitive to anisotropic magnetic noise, in other words the correlation between the noises along different directions, and how randomly-flipping spins can give rise to such anisotropic noise. 

\subsection{A. Asymmetric angle dependence of coherence time}
When the NV electron spin is in a superposition of $|0\rangle$ and $|\pm \rangle$, the coherence is affected by the transition energy fluctuation $\delta E_{\pm, 0}$ between the two states:
\begin{gather}
\delta E_{\pm, 0}(t) = \delta \vec{B} (t)\cdot (\langle \vec{S} \rangle_{\pm} - \langle \vec{S} \rangle_{0}) 
\end{gather}
Define the off-angle $\Delta \theta \equiv \theta_B - 90\degree$. Recalling Fig. 2a, b of the main text, we can write $\langle S_z \rangle _- \equiv k \Delta \theta$, $\langle S_z \rangle _+ \equiv -k \Delta \theta$, $\langle S_x \rangle _+ \approx C$ and $\langle S_x \rangle _0 \approx -C$, where $k > 0$ and $C>0$ are constants. Therefore,
\begin{gather} 
\delta E_{-0}(t) =  - \delta B_x(t) \langle S_x\rangle_{0} +  \delta B_z(t)\langle S_z\rangle_{-} =  C \cdot \delta B_x(t) + k\Delta\theta \cdot \delta B_z(t)\\
\delta E_{+0}(t) = \delta B_x(t) \left( \langle S_x\rangle_{+} - \langle S_x\rangle_{0} \right) +  \delta B_z(t) \langle S_z\rangle_{+} = 2C \cdot \delta B_x(t) - k\Delta\theta \cdot \delta B_z(t) 
\end{gather}
The energy fluctuations cause random phase accumulation and hence decoherence. Suppose $\Delta\phi$ is the phase accumulated during the free-evolution time $\tau$, then the measured spin-echo signal $\propto \cos (\Delta\phi) $. The time-averaged signal is $\langle \cos (\Delta\phi) \rangle$. Assume $\Delta\phi$ follows normal distribution centered at zero, by Taylor expansion we have $\langle \cos (\Delta\phi) \rangle = \exp(-\langle \Delta \phi ^2 \rangle /2)$, therefore the averaged spin-echo signal is determined by the variance of the phase. At a certain $\tau$, $\langle \Delta \phi ^2 \rangle \sim \langle \delta E^2 \rangle$.
\begin{gather}
\langle \Delta \phi_{-0} ^2 \rangle \sim \langle \delta E_{-0}^2 \rangle = k^2\Delta\theta^2\langle \delta B_z^2\rangle + C^2 \langle \delta B_x^2 \rangle  + 2Ck\Delta\theta \langle \delta B_x \delta B_z \rangle\\
\langle \Delta \phi_{+0} ^2 \rangle \sim \langle \delta E_{+0}^2 \rangle = k^2\Delta\theta^2\langle \delta B_z^2\rangle + 4C^2 \langle \delta B_x^2 \rangle  - 4Ck\Delta\theta \langle \delta B_x \delta B_z \rangle
\end{gather}
The last terms in Eqs. (27), (28) indicate that coherence is sensitive to the correlation between $\delta B_x$ and $\delta B_z$ noise, $\langle \delta B_x \delta B_z \rangle$. For the anisotropic noise $\langle \delta B_x \delta B_z \rangle \neq 0$, there exists an optimal $\Delta\theta_{opt} \neq 0$ which minimizes the decoherence (recalling the white dashed lines in Fig. 4c-d of the main text).  The sign of $\Delta\theta_{opt}$ depends on the sign of $\langle \delta B_x \delta B_z \rangle$ and is opposite for $|\pm\rangle$. Hence we expect to see the coherence time exhibits asymmetric angle dependence and $|\pm\rangle$ show different asymmetries.

\subsection{B. Origin of anisotropic noise in diamond}
In diamond, magnetic noise is mainly caused by nearby randomly-flipping spins in the local environment. We first explain how a single spin can give rise to such an anisotropic noise at the NV. 
\begin{figure} [h]
\centering
 \includegraphics[scale=0.87]{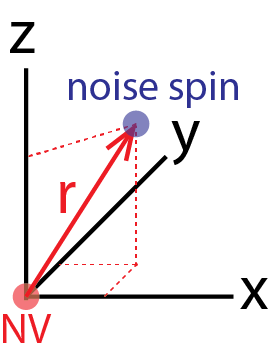}
 \caption{\label{} The noise spin is at $\vec{r}$ from the NV at the origin. The unit vector is $(u_x, u_y, u_z).$}
\end{figure}

The coupling between two magnetic dipoles is:
\begin{gather}
H_{d-d} = -\frac{\mu_0}{4\pi r^3} ( 3(\vec{m_1} \cdot \hat {\vec{r}})(\vec{m_2} \cdot \hat {\vec{r}}) -\vec{m_1} \cdot \vec{m_2}) = -\vec{m_1}\cdot \vec{B_2 (\vec{r})},
\end{gather}
where $\vec{B_2 (\vec{r})}$ is the dipolar field generated by the dipole moment $\vec{m_2}$. Decomposing all the terms into x, y and z components, we can write the magnetic field $\delta \vec{B}$ due to a nearby noise spin at $\vec{r}$ relative to the NV position (Fig. 7):
\begin{gather}
D\equiv \frac{\mu_0}{4\pi r^3} \mu_B g_e\\
\delta B_x = D[(3u_x^2-1)\hat{S_x} + 3u_x u_y \hat{S_y} + 3u_x u_z \hat{S_z}]\\
\delta B_y = D[3u_x u_y \hat{S_x} +(3u_y^2-1)\hat{S_y} +  3u_y u_z \hat{S_z}]\\
\delta B_z = D[3u_x u_z \hat{S_x} + 3u_y u_z \hat{S_y}  + (3u_z^2-1)\hat{S_z}]
\end{gather}
Here $(u_x, u_y, u_z)$ is the unit vector connecting the NV and the noise spin. $\hat{S_x}$, $\hat{S_y}$ and $\hat{S_z}$ denote the spin operators of the noise spin. Next, we calculate $\langle \delta B_{x,y,z}^2 \rangle $ using $\langle \hat{S_x}^2 \rangle = \langle \hat{S_y}^2 \rangle = \langle \hat{S_z}^2 \rangle = S^2$, $\langle  \hat{S_i} \hat{S_j} \rangle = 0 $ for $i\neq j$, and $u_x^2 + u_y^2 + u_z^2 = 1$. It follows that
\begin{gather}
\langle \delta B_x^2 \rangle = D^2  S^2 (3u_x^2+1) \\
\langle \delta B_z^2 \rangle = D^2  S^2 (3u_z^2+1) \\
\langle \delta B_x \delta B_z \rangle = D^2  S^2 (3u_x u_z)
\end{gather}
According to Eq. (36), $\delta B_x$ and $\delta B_z$ are correlated as long as $u_x\neq0$ and $u_z\neq0$. Recalling Eqs. (27), (28), we expect to see the NV coherence shows angle asymmetry. 

The above is the result from a single noise spin. On the other hand, for a large ensemble of randomly distributed noise spins, the sum of their individual terms $\langle \delta B_x \delta B_z \rangle$ is likely be close to zero and the noise then becomes isotropic. The asymmetry observed in our experiment indicates that the NV magnetic noise is dominated by one or a few spins nearby, and the summed effect is anisotropic. 

Finally, we point out the difference between the real noise generated by randomly-flipping spins in diamond and a fully anisotropic noise fluctuating perfectly along a straight line, such that $\delta B_x$ and $\delta B_z$ are proportional to each other: $\delta B_x = t \delta B_z (t\neq0)$. In the latter case, we have
\begin{gather}
\langle \delta B_x^2 \rangle = t^2 \langle \delta B_z^2 \rangle = t\langle \delta B_x \delta B_z \rangle
\end{gather}
Plugging Eq. (37) into Eqs. (27), (28), there exists an optimal $\Delta \theta_{opt}$, at which the noise is completely suppressed, i.e. $\langle \delta E_{\pm0}^2 \rangle = 0$ (see Fig. 4c-d of the main text). However, for real noise from randomly-flipping spins (Eqs. (31) - (33)), at the optimal $\Delta \theta_{opt}$, the noise coupling is maximally reduced but not zero.

\section{IV. Simulation details}
We simulated the NV spin-echo measurement using the QuTip package (Quantum Toolbox in Python) \cite{Johansson2012, Johansson2013}. We first describe setting up the simulation in the rotating frame, and then incorporating noise into the system.

\subsection{A. Rotating frame simulation}
The rotating frame is defined by the electron spin energy eigenstates. $\hat{U}_{-0}$ ($\hat{U}_{+0}$) represent the rotating frame transformation operators for simulating the spin-echo between $|0\rangle$ and $|-\rangle$ (or $|0\rangle$ and $|+\rangle$):
\begin{eqnarray}
\hat{U}_{-0} = e^{-iE_{-0}t|-\rangle\langle -|}\\
\hat{U}_{+0} = e^{-iE_{+0}t|+\rangle\langle +|}
\end{eqnarray}
The electron spin (S=1) is prepared in a superposition of $|0\rangle$ and $|-\rangle$: $|\psi_0\rangle = \frac{1}{\sqrt{2}}(|0\rangle + |-\rangle)$, and the $^{15}N$ nuclear spin (I = 1/2) is in an uncontrolled mixed state: $\rho_{N0} = \begin{bmatrix} 0.5 & 0 \\ 0 & 0.5 \end{bmatrix}$. The full NV spin Hamiltonian is:
\begin{eqnarray}
H_{gs} = H_{gs,e} + H_{gs,n} = D_{gs}S_z^2 + \gamma_B \vec{B} \cdot \vec{S} + \vec{I}\cdot\vec{A}\cdot\vec{S} + \gamma_N \vec{B} \cdot \vec{I} 
\end{eqnarray}
Couplings to strains or electric fields are weak and not considered here. The rotating frame Hamiltonian is:
\begin{eqnarray}
H_{rot} = \hat{U}_{rot}^{-1} H_{gs} \hat{U}_{rot} - i \hat{U}_{rot}^{-1} \frac{d\hat{U}_{rot}}{dt},
\end{eqnarray}
where $\hat{U}_{rot} = \hat{U}_{-0} $ or  $\hat{U}_{+0}$.
The $\pi$ pulse operator in the rotating frame is:
\begin{eqnarray}
R_{\pi} = |0\rangle \langle \pm| +  |\pm\rangle \langle 0|
\end{eqnarray}
The simulation starts from the initial density matrix $\rho_0= |\psi_{0}\rangle \langle \psi_{0}| \otimes \rho_{N0}$, and its time evolution follows the rotating frame Hamiltonian $H_{rot}$ with a $\pi$ pulse in the middle. The spin-echo signal calculates the overlap between the electron spin final and initial states, tracing out the nuclear spin state. 

\subsection{B. Simulation with noise}
Collapse operators were defined to simulate the decoherence processes, and time evolution is governed by the Lindblad master equation in QuTip. To simulate a magnetic noise $\delta \vec{B}_{\gamma}(t)$ along a certain direction $\hat{\gamma}$, which couples to the electron spin operators $S_{\hat{\gamma}}$, the collapse operator is defined as $C \equiv \sqrt{\Gamma} S_{\hat{\gamma}}$, where $S_{\hat{\gamma}}$ is a linear combination of $S_x$, $S_y$ and $S_z$, and $\Gamma$ is the coupling strength. Multiple collapse operators can be defined if more than one noise exist in the environment. 

To simulate a noise fluctuating along a straight line at $-45\degree / +135\degree $ from $+\hat{x}$ (Fig. 4e of the main text),  we define the collapse operator:
\begin{eqnarray}
C = \sqrt{\Gamma} \left(S_x - S_z\right),
\end{eqnarray}
where $S_x$ and $S_z$ are the spin operators transformed to the rotating frame. However, if the noise is isotropic, such that the $\hat{x}$ and $\hat{z}$ components are uncorrelated, we need to define a list of two separate collapse operators: $[\sqrt{\Gamma} S_x, \sqrt{\Gamma} S_z ]$.
Fig. 4c-d of the main text visualize the simulation results in the presence of such noise.

Here we point out two differences between the simulation and experiment. Firstly, the simulated noise in QuTip is Markovian noise (white noise), which has infinitely short correlation time. However, the real noise in our experiment has specific noise spectrum and finite correlation time (recalling that dynamical decoupling can successfully extend the coherence). This inconsistency can cause an incorrect estimate of the coherence time and the exact shape of the coherence decay as a function of $\tau$. Secondly, we only included a fully anisotropic noise in the simulation, however in real experiments, multiple noise sources coexist in the sample and the summed effect is partially anisotropic. In addition to magnetic field noise, there also exist electric field noise, temperature fluctuations, etc. which were not taken into account. 

 \begin{figure}[h]
 \includegraphics[scale=0.33]{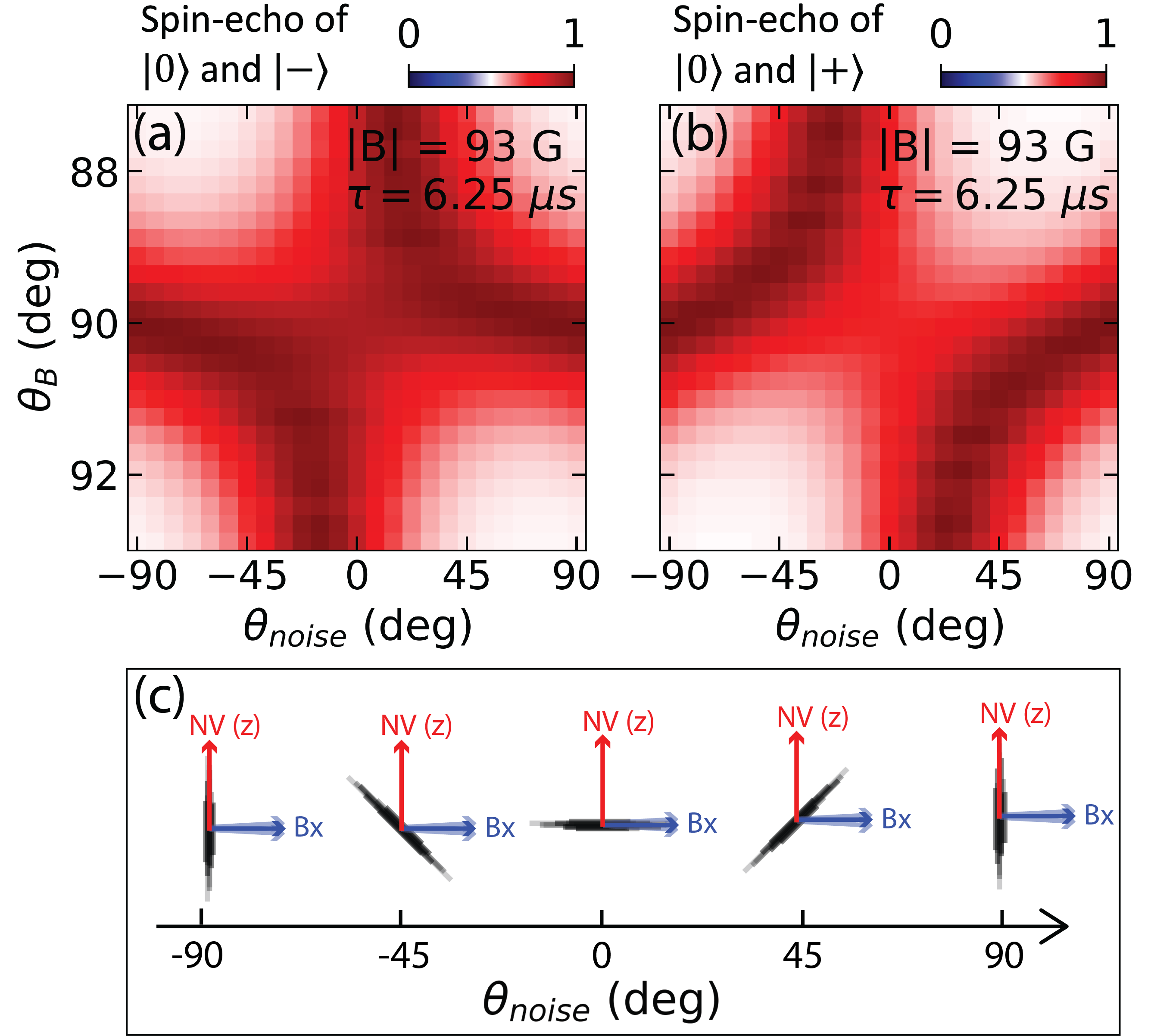}
  \caption{\label{} (a)(b) Asymmetric angle dependence of the NV coherence time as the noise direction rotates in the XZ plane. Spin-echo simulation is done at a fixed free evolution time $\tau = 6.25$ $\mu$s under the bias field $|B|$ = 93 G. Colorbar represents the spin-echo amplitude. $\theta_{noise}$ is the angle between the noise and $\hat{x}$. (c) Schematics of the noise direction (black lines) relative to the NV axis and the bias magnetic field direction projected in the XZ plane.}
 \end{figure}

To better illustrate that this provides a way to characterize the noise direction, we performed spin-echo simulations at fixed $\tau=6.25$ $\mu$s as a function of the noise direction. As shown in Fig. 8, the angle asymmetry significantly changes as the noise direction rotates in the XZ plane. 
\nocite{*}
\bibliography{supplementary_reference}